\RequirePackage{rotating}
\documentclass[preprint]{imsart}

\RequirePackage{amsthm,amsmath,amsfonts,amssymb,mathtools,bm,mathabx,tikz}
\RequirePackage[authoryear]{natbib}
\RequirePackage[colorlinks,citecolor=blue,urlcolor=blue]{hyperref}
\RequirePackage{graphicx}
\usepackage[boxed, ruled, lined, linesnumbered]{algorithm2e}
\usepackage[nameinlink]{cleveref}
\usepackage{multirow}
\usepackage{color}
\usepackage{listings}
\startlocaldefs
\numberwithin{equation}{section}
\theoremstyle{plain}
\newtheorem{thm}{Theorem}
\numberwithin{thm}{section}

\newtheorem{lem}{Lemma}
\numberwithin{lem}{section}

\newtheorem{prop}{Proposition}
\numberwithin{prop}{section}

\newtheorem{cor}{Corollary}
\numberwithin{cor}{section}

\theoremstyle{remark}
\newtheorem{ass}{Assumption}

\numberwithin{defn}{section}

\newtheorem{rem}{Remark}
\numberwithin{rem}{section}

\newenvironment{ass*}
 {\expandafter\def\expandafter\theass\expandafter{\theass*}\ass}
 {\endass}

\Crefname{ass}{Assumption}{Assumptions}
\Crefname{prop}{Proposition}{Propositions}
\Crefname{section}{Section}{Sections}
\Crefname{appendix}{Appendix}{Appendices}
\Crefname{cor}{Corollary}{Corollaries}

\newcommand{\R}{\mathbb{R}}

\DeclareMathOperator*{\argmin}{arg min}
\DeclareMathOperator*{\argmax}{arg max}
\DeclareMathOperator{\Vector}{vec}
\DeclareMathOperator{\Vech}{vech}
\DeclareMathOperator{\tr}{tr}
\DeclareMathOperator{\offvec}{offvec}
\DeclareMathOperator{\blkdiag}{blkdiag}
\DeclareMathOperator{\SNR}{SNR}
\DeclareMathOperator{\mat}{mat}
\DeclareMathOperator{\cov}{Cov}
\DeclareMathOperator{\var}{Var}
\DeclareMathOperator{\diag}{diag}
\newcommand{\rank}{\operatorname{rk}} 
\endlocaldefs

\begin{document}

\begin{frontmatter}
\title{Testing Simultaneous Diagonalizability}
\runtitle{Testing Simultaneous Diagonalizability}

\begin{aug}
\author[A]{\fnms{Yuchen}~\snm{Xu}\ead[label=e1]{yx439@cornell.edu}},
\author[A]{\fnms{Marie-Christine}~\snm{D\"uker}\ead[label=e2]{md2224@cornell.edu}}
\and
\author[A]{\fnms{David S.}~\snm{Matteson}\ead[label=e3]{matteson@cornell.edu}}
\address[A]{Department of Statistics and Data Science, Cornell University\printead[presep={,\ }]{e1,e2,e3}}
\end{aug}

\begin{abstract}
This paper proposes novel methods to test for simultaneous diagonalization of possibly asymmetric matrices. Motivated by various applications, a two-sample test as well as a generalization for multiple matrices are proposed. A partial version of the test is also studied to check whether a partial set of eigenvectors is shared across samples. Additionally, a novel algorithm for the considered testing methods is introduced. Simulation studies demonstrate favorable performance for all designs. 
Finally, the theoretical results are utilized to decouple multiple vector autoregression models into univariate time series, and to test for the same stationary distribution in recurrent Markov chains.
These applications are demonstrated using macroeconomic indices of 8 countries and streamflow data, respectively.
\end{abstract}



\end{frontmatter}


\section{Introduction}\label{sec:intro}
Understanding the eigenvectors and the eigenspace of matrix-valued objects is known to be of fundamental interest in various disciplines including statistics, machine learning, and computer science.
Knowledge about the eigenvectors and the eigenspace is particularly valuable in principal component analysis (PCA) 
\citep{nadler2008finite,cai2013sparse,koltchinskii2017new}
, covariance matrix estimation \citep{fan2013large,fan2015estimation,fan2018eigenvector}, spectral clustering \citep{von2007tutorial,rohe2011spectral,lei2015}, and network or graph theory \citep{tang2018,paul2020spectral}.
Often times it provides information for dimension reduction and clustering procedures.

This paper develops statistical tests and algorithms to check whether a set of square matrices can be diagonalized simultaneously. We are particularly interested in asymmetric square matrices with more general and flexible structural assumptions compared to symmetric ones like covariance matrices. Our work proceed from two-sample tests to multi-sample tests, and finally extends into partial cases where only a subset of eigenvectors is of interest.
Besides providing the theoretical foundation and introducing practical algorithms, we motivate the usefulness of our results in several examples.

Our setting is as follows.
Suppose we have a sequence of deterministic matrices $\{M_i\}_{i = 1}^p \subset \mathbb{R}^{d \times d}$. Then the hypothesis testing problem we are interested in can be expressed as: the null hypothesis is $H_0$: $\{M_i\}_{i = 1}^p$ can be jointly diagonalized, or equivalently, 
\begin{equation}\label{hyp:all}
    H_0: ~ \exists V \in \mathbb{R}^{d \times d}, ~ s.t. ~ M_i V = V D_i,\ \forall i = 1, \dots, p, ~ D_i \in \mathcal{M}_{\diag}(d),
\end{equation}
where $\mathcal{M}_{\diag}(d)$ denotes the set of diagonal matrices in $\mathbb{R}^{d \times d}$.
A modification of the problem is to discuss whether a set of matrices shares a partial set of eigenvectors. The null hypothesis is then expressed as $H_0^*$: $\{M_i\}_{i = 1}^p$ share $k$ left eigenvectors ($k<d$), or equivalently, 
\begin{equation} \label{hyp:part}
    H_0^*: ~ \exists V = (v_1, \dots, v_k) \in \mathbb{R}^{d \times k} \text{ full rank},\ s.t. ~ M_i V = V D_i, ~ D_i \in \mathcal{M}_{\diag}(k).
\end{equation}

In a series of contributions \cite{Flury84,FluryAsymptotic,Flury86} introduced common principal component analysis (CPCA) that deals with the test and calculation of simultaneous factorization among different samples of positive definite symmetric matrices. 
Under the same assumptions, \cite{schott} developed the terminology partial CPCA (PCPCA) with verified test methods for partially identical eigenvectors. 
In contrast, we do not need to impose any structural assumptions on the pool of matrices like symmetry or positive semidefinitenss which are naturally provided by considering covariance matrices.

Related to simultaneous diagonalization, previous studies mainly focused on testing whether the eigenvectors or eigenspaces of the population covariance matrix are equal to some given ones; see \cite{tyler,koltchinskii2017new,silin2018bayesian,naumov2019bootstrap,silin2020hypothesis}. \cite{Schwartz} studied some related statistical tests about eigenvalues and eigenvectors of Gaussian random symmetric matrices with some pre-fixed algebraic restrictions.
Especially, as stated in \cite{Schwartz}, the test for equality of eigenvectors with unknown eigenvalues between two sets of samples is rather difficult since no closed forms of estimations are available.

From a computational perspective, optimization routines for symmetric matrices were proposed by \cite{Fuji, Ghazi, Gira}. For general asymmetric matrices, some previous ideas, like '\textit{sh-rt}' by \cite{shrt}, '\textit{JUST}' by \cite{just}, '\textit{JDTM}' by \cite{jdtm}, and '\textit{(W)JDTE}' by \cite{andre} are shown to be numerically effective and ready for implementation. \cite{colombo, tensor} focused on the joint Schur-decomposition and provided theoretical properties of their proposed algorithms. Since joint Schur-decomposability fails to be a sufficient condition for simultaneous diagonalization, our work expands those ideas and provides an algorithm which estimates partially common eigenvectors across samples. 

Possibly asymmetric matrix-valued statistics are broadly utilized in estimating the mean of random matrices, the adjacency matrices of weighted directed graphs, the coefficient matrices in linear regressions, factor models and vector autoregression (VAR) models, and transition probability matrices. However, most of the analysis has focused on studying the eigenvalues of those statistics. Analyses of eigenvalues include reduced rank estimation \citep{robin:2000tests, Kleibergen2006:Generalized, donald:2007rank}, testing for cointegration 
\citep{Engle1987:Co,johansen91:Est,vogelsang2001unit,zhang2019identifying}
and the eigenvalues of adjacency matrices \citep{restrepo2007approximating,paul2020spectral}. In contrast, our applications (see \autoref{subs:applic} below) give a new perspective on the usefulness of studying the eigenvectors in various models.

The literature review shows that existing work is based on covariance matrices which are surely diagonalizable with orthogonal eigenvectors. 
The eigenstructure problem lacks analysis in some more general cases like asymmetric matrices in particular. The breakthrough point of our work is to design and validate efficient diagonalization test methods for possibly asymmetric matrices.
Due to non-linearity of eigenproperties and the lack of closed eigensolutions,
our investigation about the random eigenstructures with less restricted conditions is algebraically difficult, and our exploration is novel.

\subsection{Applications} \label{subs:applic}
From a statistical perspective, joint diagonalizability provides valuable information.
Suppose one fails to reject the null hypothesis of common eigenvectors, then, it is reasonable to only analyze the eigenvalues, which reduces the problem's complexity significantly. 
We will illustrate the usefulness of our results with two relevant examples, namely the coefficient matrices in VAR models and the transition matrices of Markov chains.

The coefficient matrices in VAR models appear to be general matrices without restrictions like symmetry. 
For the same multivariate time series of comparable objects, simple VAR models of order one may share common components in regression which can be verified by a joint eigendecomposition of coefficient matrices. With successful verification of simultaneous diagonalizability, one can decouple multivariate time series into multiple univariate ones and conduct comparison conveniently; see \Cref{se:VAR} for more details.

Another motivating example for our tests are the transition matrices of Markov chains which are usually asymmetric. Furthermore, the leading left eigenvector of a transition matrix corresponds to eigenvalue one and represents the stationary distribution of the chain. Testing the equality of the leading eigenvectors of the transition matrices from multiple Markov chains gives information whether these chains share similar properties though differing in their transition dynamics. For instance, the Markov chains of the same object but with different time resolutions might exhibit a common stationary distribution; see also \autoref{se:Markov} for more details on this application.


\subsection{Organization}
The rest of this paper is organized as follows. \autoref{sec:prel} establishes notation and gives some preliminary results.
Our main work starts in \autoref{sec:commute} from a two-sample test. In addition to simultaneously conducting two-sample tests pairwise for multi-sample cases, we design our test method based on the pooled estimator of common eigenvectors; see \autoref{sec:multi}. In \autoref{sec:part} we further extend our results to a partial version with a novel algorithm to estimate the subset of eigenvectors that is shared across samples. We also briefly show the compatibility of our test methods with the symmetric setting in \autoref{sec:ext}. In \Cref{sec:simu,,sec:app} we conduct a simulation study and experiment with some real data examples, respectively. The supplementary material provides some empirical results in \autoref{se:compl} complementary to the numerical analysis presented in the main paper.
Finally, the proofs of the theoretical results in \Cref{sec:twosample,sec:multi,sec:part} can be found in the supplementary material in \Cref{app:profs,,app:est}.

\section{Preliminaries} \label{sec:prel}
\subsection{Notation} \label{subsec:not}
Throughout this paper, $p$ is the number of matrices to be tested, $n$ denotes the sample size for estimation, and $d$ is the dimension of the square matrices. Notation $\xrightarrow{\mathcal{D}}$ represents convergence in distribution, $\xrightarrow{\mathcal{P}}$ convergence in probability, and $\stackrel{\mathcal{D}}{\approx}$ an approximation of random distributions. The operator $\otimes$ denotes the Kronecker product between two matrices.
For a matrix $A$, the operator $\rank(A)$ denotes the rank of $A$, $\Vector(A)$ transforms $A$ into a vector form by stacking all its columns, and $A^+$ is the Moore-Penrose general inverse of $A$. For a square matrix $A$ of dimension $d$, we write $\tr(A)$ for the trace function of $A$ and the operator $\mat_d(\cdot)$ is the inverse of $\Vector(\cdot)$ such that $\mat_d\big(\Vector(A)\big) = A$.
The matrix $I_d$ represents the $d$-dimensional identity matrix, and the function $\blkdiag(\{X_i\}_{i=1}^p)$ returns a block-diagonal matrix with the sub-matrices on the diagonal to be the input list of matrices $\{X_i\}_{i=1}^p$. We write $\mathcal{N}(\mu, \Sigma)$ for the multivariate normal distribution with mean vector $\mu$ and covariance matrix $\Sigma$, $\chi^2(k)$ for the chi-squared distribution with $k$ degrees of freedom, and $\gamma(\alpha, \beta)$ for the Gamma distribution with shape parameter $\alpha$ and rate parameter $\beta$.

\subsection{Assumptions} \label{subsec:ass}
In this section, we give the required assumptions for future proof of the asymptotic results for our proposed test statistics. 

\begin{ass}\label{ass:normal}
The deterministic matrices $\{M_i\}_{i = 1}^p$ can be estimated by mutually independent estimators $\{A_{i,n}\}_{i = 1}^p$ from $n$ samples, satisfying,
\begin{equation}\label{eq:normal}
    c(n)\Vector(A_{i,n} - M_i) \xrightarrow{\mathcal{D}} \mathcal{N}(0, \Sigma_i) \hspace{0.2cm} \text{ for } \hspace{0.2cm} i=1,\dots,p,
\end{equation}
with $c(n) \to \infty$ as $n \to \infty$.
\end{ass}

\begin{ass}\label{ass:covConsistent}
The limiting covariance matrices $\Sigma_i$ in \eqref{eq:normal} can be estimated consistently by $\widehat{\Sigma}_{i,n}$, that is,
\begin{equation}\label{eq:sigma.rate}
    \widehat{\Sigma}_{i,n} - \Sigma_i = O_{\mathcal{P}}\big(\sigma(n)\big)
     \hspace{0.2cm} \text{ for } \hspace{0.2cm} i=1,\dots,p,
\end{equation}
with rate $\sigma(n) \to 0$ as $n \to \infty$.
\end{ass}

Note that the estimators $A_{i,n}$ and $\widehat{\Sigma}_{i,n}$ also depend on the sample size $n$, but for notational simplicity we will omit it in later expressions and simplify as $A_i = A_{i,n}$ and $\widehat{\Sigma}_i = \widehat{
\Sigma}_{i,n}$. In addition, we assume the sequence $c(n)$ to be the same for all $i = 1, \dots, p$, in the following analysis but the extension to the general case is straightforward. Furthermore, we assume the following:
\begin{ass}\label{ass:distinct}
Each $M_i$, $i=1,\dots,p$ is diagonalizable with real eigenvalues.
\end{ass}


The theoretical results of this paper require no a priori assumptions on the rank of the limiting covariance matrices in \autoref{ass:normal}. In particular, the matrices $\Sigma_{i}$, $i = 1, \dots, p$, in \eqref{eq:normal} may be less than full rank. On one hand, this allows for flexibility in the choice of estimators $A_{i,n}$, on the other hand it elevates the difficulties in deriving asymptotic results for our test statistics. In particular, we aim to give tractable versions of our test statistics in the sense that the covariance matrices can be estimated. While \autoref{ass:covConsistent} ensures the existence of a consistent estimator, one still has to address a potential singularity. We refer to \autoref{subsec:Cov} for a discussion and workaround.

\subsection{Covariance estimation} \label{subsec:Cov}
Our test statistics involve the inverses and ranks of the limiting covariance matrices in \autoref{ass:normal}. Ideally, given exact covariance matrices, we can use the so-called Moore-Penrose inverse to address possible singularity. In order to make the statistics tractable in practice, we need to instead incorporate their consistent estimates as given by \autoref{ass:covConsistent}. However, neither the Moore-Penrose inverse nor the rank of a matrix are continuous. 

To circumvent those issues, we introduce the so-called truncated singular value decomposition following \cite{lutkepohl1997modified}.
For an arbitrary matrix $\Psi \in \mathbb{R}^{a \times b}$ with $\tau = \min(a, b) > 1$, its singular value decomposition (SVD) is given by $\Psi = U \Pi W'$ with orthogonal singular vectors $U$ and $W$, and non-increasing, non-negative singular values $\Pi = \diag(\varpi_1, \dots, \varpi_\tau)$.
We define its truncated singular value decomposition with respect to a threshold $\varepsilon \geq 0$ as 
\begin{equation} \label{eq:truSVD}
\Psi(\varepsilon) = U \Pi(\varepsilon) W'
\hspace{0.2cm}
\text{ with }
\hspace{0.2cm}
\Pi(\varepsilon) = \diag(\varpi_1 \mathbb{I}(\varpi_1 > \varepsilon), \dots, \varpi_\tau \mathbb{I}(\varpi_\tau > \varepsilon))
\end{equation}
with indicator function $\mathbb{I}(\cdot)$. In addition, we denote the Moore-Penrose general inverse as $\Psi^+(\varepsilon) = \big(\Psi(\varepsilon)\big)^+$ and the rank function $\rank(\Psi; \varepsilon) = \rank\big(\Psi(\varepsilon)\big)$. 

With the help of the truncated SVD \eqref{eq:truSVD}, we introduce the following lemma that gives consistent estimates for the Moore-Penrose inverse and the rank of a consistently estimated matrix.
\begin{lem}\label{lem:ginverse}
Assume $\widehat{\Sigma}$ is a consistent estimator of a positive semidefinite matrix $\Sigma \in \mathbb{R}^{\tau \times \tau}$, $\varepsilon > 0$ is a constant and not an eigenvalue of $\Sigma$. Then, with $\Sigma(\varepsilon)$ and $\widehat{\Sigma}(\varepsilon)$ defined to be the corresponding truncated SVDs \eqref{eq:truSVD}, 
\begin{equation}\label{eq:wald.cmt}
    \widehat{\Sigma}(\varepsilon) \xrightarrow{\mathcal{P}} \Sigma(\varepsilon), ~ \widehat{\Sigma}^+(\varepsilon) \xrightarrow{\mathcal{P}} \Sigma^+(\varepsilon), ~ \rank(\widehat{\Sigma}; \varepsilon) \xrightarrow{\mathcal{P}} \rank(\Sigma; \varepsilon).
\end{equation}
\end{lem}

The proof of \autoref{lem:ginverse} can be found in \autoref{app:profs}.
Results in the manner of \autoref{lem:ginverse} are used to circumvent singularity issues which usually occur under the usage of Wald type tests; see 
\cite{hadi1990note, ratsimalahelo2001rank}.
The following remark comments on the choice of the threshold $\varepsilon$ in \eqref{eq:wald.cmt}.

\begin{rem}\label{rem:epsilon.order}
With the additional \autoref{ass:covConsistent} that $\widehat{\Sigma}$ converges with rate $O_\mathcal{P}\big(\sigma(n)\big)$, the threshold $\varepsilon$ that satisfies $\varepsilon = o(1)$ and $\sigma(n)/\varepsilon = o(1)$ as $n \to \infty$ can be chosen to optimize the accuracy of the generalized inverse and rank estimators. See p.\ 320 in \cite{lutkepohl1997modified} for a discussion on the choice of $\varepsilon$.
\end{rem}

\section{Two-sample test}\label{sec:twosample}
We start from a two-sample test $p = 2$. In \autoref{sec:commute}, we design a test statistic based on the commutator of the two matrices under consideration. In \autoref{sec:llrt}, the log-likelihood ratio test is adjusted and in \autoref{sec:llrintro} a test statistic for the hypothesis testing problem \eqref{hyp:all} is introduced, along with its sensitivity analysis in \autoref{sec:error}. We conclude the section with \autoref{sec:sum}.

\subsection{Commutator-based test}\label{sec:commute}
Under \autoref{ass:distinct}, matrices commute if and only if they can be diagonalized simultaneously; see Theorem 1.3.12 in \cite{Horn}. Hence, one intuitive idea to measure how far $M_1$ and $M_2$ are from being commutable is, to calculate some form of metric of their commutator $[M_1, M_2] := M_1 M_2-M_2 M_1$. The following proposition introduces a statistic to test the hypothesis \eqref{hyp:all} and provides its asymptotic behavior. 
\begin{prop}\label{thm:comm}
Suppose \Cref{ass:normal,ass:distinct} are satisfied and denote $\bm{\eta}_n = \Vector[A_1, A_2]$. Then, under $H_0$ in \eqref{hyp:all},
\begin{equation}\label{eq:comm.asym}
    c(n) \bm{\eta}_n \xrightarrow{\mathcal{D}} \mathcal{N}(0, \Sigma_\eta)
\end{equation}
with $\Sigma_\eta = \Sigma_{1,2} + \Sigma_{2,1}$, where $\Sigma_{k, \ell} = \Lambda(M_\ell) \Sigma_k \Lambda'(M_\ell)$ for $k \ne \ell$, and $\Lambda(X) = I_d \otimes X - X' \otimes I_d$ is a function in $X \in \R^{d \times d}.$ Then,
\begin{equation}\label{eqn:gamma2}
    \Gamma_1 := c^2(n) \bm{\eta}_n' \Sigma_\eta^+ \bm{\eta}_n \xrightarrow{\mathcal{D}} \chi^2(r_1),
\end{equation}
where $r_1$ is the rank of $\Sigma_\eta$.
\end{prop}

The proof of \autoref{thm:comm} can be found in \autoref{app:profs}. In order to make \autoref{thm:comm} tractable in practice, we can obtain a consistent estimator $\widehat{\Sigma}_\eta$ by substituting $\Sigma_{i}, M_{i}$, $i=1,2$, in the expression of $\Sigma_\eta$ with $A_{i},\widehat{\Sigma}_{i}$, $i=1,2$, in \Cref{ass:normal,,ass:covConsistent}, respectively. The consistency is verified by the continuous mapping theorem and \Cref{ass:normal,ass:covConsistent} as
$$
\widehat{\Sigma}_\eta - \Sigma_\eta = \Lambda(A_2) \widehat{\Sigma}_1 \Lambda'(A_2) + \Lambda(A_1) \widehat{\Sigma}_2 \Lambda'(A_1) - \Sigma_\eta \xrightarrow{\mathcal{P}} 0.
$$
Note that $\Sigma_\eta$ and $\widehat{\Sigma}_\eta$ are both singular matrices as there exists at least one non-trivial vector $\bm{v} = \Vector(I_d)$ such that $\Sigma_\eta \bm{v} = \widehat{\Sigma}_\eta \bm{v} = 0$, since
$$\Lambda(X)\bm{v} = (I_d \otimes X)\bm{v} - (X' \otimes I_d)\bm{v} =0,$$
where the last equality follows by Theorem 2 in \cite{magnus2019matrix}, p.\ 35.
Hence $r_1$ and $\hat{r}_1$, the ranks of $\Sigma_\eta$ and $\widehat{\Sigma}_\eta$ respectively, are always less than $d^2$. Due to the singularity issue
in \eqref{eqn:gamma2}, we propose to use the truncated version \eqref{eq:wald.cmt} of $\widehat{\Sigma}_\eta$.

\begin{prop}\label{thm:comm.Est}
Suppose \Cref{ass:normal,,ass:covConsistent,,ass:distinct} are satisfied. Then, for a given threshold $\varepsilon > 0$ that is not an eigenvalue of $\Sigma_{\eta}$ (defined through \eqref{eq:comm.asym}),
the test statistic is defined as and satisfies
\begin{equation}\label{eq:gamma2.est}
    \Gamma_1^\#(\varepsilon) := c^2(n) \bm{\eta}_n' \widehat{\Sigma}_\eta^+(\varepsilon) \bm{\eta}_n \xrightarrow{\mathcal{D}} \chi^2(\widehat{r}_1(\varepsilon)),
\end{equation}
where $\widehat{r}_1(\varepsilon) = \rank(\widehat{\Sigma}_\eta; \varepsilon)$.
\end{prop}

\subsection{Log-likelihood Ratio (LLR) test}
\label{sec:llrt}

In addition to the commutator based test, we propose an alternative test based on the likelihood ratio test framework in this section.

Before we introduce the test statistic we state an assumption which is slightly stronger than \autoref{ass:distinct}.

\setcounter{ass}{2}
\begin{ass*}\label{ass:distinct2}
Each $M_i$, $i=1,\dots,p$, has $d$ distinct non-zero real eigenvalues.
\end{ass*}

According to the assumed asymptotic normality in \autoref{ass:normal}, we introduce the log-likelihood type function for the estimators $A_1$ and $A_2$ as
\begin{equation}\label{eqn:loglike}
    L(M_1, M_2) := - \sum_{i = 1}^2 \Vector(A_i - M_i)' \Sigma_i^+ \Vector(A_i - M_i).
\end{equation}
It is then possible to obtain the supremum of $L(M_1, M_2)$ within the parameter spaces $H_0$ and $H_0 \cup H_1$, respectively, as
\begin{equation}\label{eqn:llrMLE}
    \widetilde{L}_0 := \sup_{(M_1, M_2) \in H_0} L(M_1, M_2), ~ ~ \widetilde{L}_1 := \sup_{(M_1, M_2) \in H_0 \cup H_1} L(M_1, M_2).
\end{equation}
In particular, we introduce a new version of M-estimators for $(M_1, M_2)$ under the null hypothesis $H_0$ as
\begin{equation}
    (\widehat{A}_1, \widehat{A}_2) = \argmax_{(M_1, M_2) \in H_0} L(M_1, M_2),\label{eqn:Mest}
\end{equation}
and the design of the ratio-test statistic can be given by
$
\Gamma_2 ~ \propto ~ - (\widetilde{L}_0 - \widetilde{L}_1) = - \widetilde{L}_0.
$

Indeed the estimators $\widehat{A}_1$ and $\widehat{A}_2$ in \eqref{eqn:Mest} can be explicitly computed given \autoref{ass:distinct2}. We introduce the following proposition and prove it in \autoref{app:profs}.

\begin{prop}\label{thm:ahat}
Suppose \autoref{ass:distinct2}.
Then, the optimizer $(\widehat{A}_1,\widehat{A}_2)$ that maximizes \eqref{eqn:loglike} under $H_0$ is given by
\begin{equation}\label{eqn:Ahat}
    \begin{aligned}
        & \Vector(\widehat{A}_1) = P_2 (P_2' \Sigma_1^+ P_2)^+ P_2' \Sigma_1^+ \Vector(A_1),\
        & \Vector(\widehat{A}_2) = P_1 (P_1' \Sigma_2^+ P_1)^+ P_1' \Sigma_2^+ \Vector(A_2),
    \end{aligned}
\end{equation}
with $P_i$ for $i = 1, 2$ generated from either of the two following setups:\begin{itemize}
    \item Polynomial basis: with $M_i^j$ the $j$-th power of $M_i$, for $i = 1, 2$,
    \begin{equation}\label{eqn:polyP}
        P_i = (\Vector(M_i^1)/\|M_i^1\|_F, \dots, \Vector(M_i^{d})/\|M_i^d\|_F).
    \end{equation}
    \item Eigenvector basis: with $V = (v_1, \dots, v_d) \in \mathbb{R}^{d \times d}$ the common eigenvectors of $M_1$ and $M_2$, $U = (u_1, \dots, u_d)' = V^{-1}$,
    \begin{equation}\label{eqn:eigvP}
        P_1 = P_2 = (\Vector(v_1 u_1'), \dots, \Vector(v_d u_d')).
    \end{equation}
\end{itemize}
\end{prop}

\subsubsection{LLR test statistic}\label{sec:llrintro}
In this section we introduce the LLR test statistic and provide its asymptotic behavior.
Under ideal conditions such that the true matrices $M_1$ and $M_2$ are known, we introduce the LLR test statistic
\begin{equation}\label{eqn:gamma1}
\begin{aligned}
    \Gamma_2 := c^2(n) \Big[ & \Vector(A_1 - \widehat{A}_1)' \Sigma_1^+ \Vector(A_1 - \widehat{A}_1) + \Vector(A_2 - \widehat{A}_2)' \Sigma_2^+ \Vector(A_2 - \widehat{A}_2)\Big]\\
    = c^2(n) \Big[ & \Vector (A_1)' Q_{1,2} \Vector(A_1) + \Vector(A_2)' Q_{2,1} \Vector(A_2)\Big],
\end{aligned}
\end{equation}
where $Q_{k,\ell} = \Sigma_k^+ - \Sigma_k^+ P_\ell (P_\ell' \Sigma_k^+ P_\ell)^+ P_\ell' \Sigma_k^+$ for $k, \ell = 1, 2$ and $k \ne \ell$,
and present its asymptotic behavior in the following proposition.
\begin{prop}[LLR test statistic]\label{thm:LLR}
Suppose \Cref{ass:normal,ass:distinct2} are satisfied. Then, under $H_0$ in \eqref{hyp:all}, the test statistic $\Gamma_2$ in \eqref{eqn:gamma1} satisfies
\begin{equation}\label{eqn:projTest}
    \Gamma_2 \xrightarrow{\mathcal{D}} \chi^2(r_2),
\end{equation}
where $r_2 = r_{1,2} + r_{2,1}$ and $r_{k,\ell} = \rank(\Sigma_k) - \rank(P_{\ell}' \Sigma_k^+ P_{\ell})$ for $k, \ell = 1,2$, and $k \neq \ell$.
\end{prop}


Note that when $\Sigma_1$ and $\Sigma_2$ are non-singular, $r_2 = 2d^2 - 2d$.
With our loose constraints on the covariance matrices, we may encounter singularity issues when computing \eqref{eqn:gamma1} with $\Sigma_1^+$ and $\Sigma_2^+$. To have a tractable version of \autoref{thm:LLR} with respect to the limiting covariance matrices, we propose to use the truncated version \eqref{eq:wald.cmt}. Note that the generalized inverse of $P_\ell' \Sigma_k^+ P_\ell$ is a part of a projection matrix hence will not have the same discontinuity concerns.


\begin{prop}\label{thm:LLRest}
Suppose \Cref{ass:normal,,ass:covConsistent,,ass:distinct2} are satisfied. Let $\varepsilon > 0$ be a threshold that is not an eigenvalue of $\Sigma_1$ and $\Sigma_2$. Define the test statistic
\begin{equation}\label{eqn:gamma1.Est}
    \Gamma_2^\#(\varepsilon) := c^2(n) \Big[\Vector (A_1)' \widehat{Q}_{1,2}[\varepsilon] \Vector(A_1) + \Vector(A_2)' \widehat{Q}_{2,1}[\varepsilon] \Vector(A_2)\Big]
\end{equation}
with
\begin{equation} \label{eq:Qkl}
\widehat{Q}_{k,\ell}[\varepsilon] := \widehat{\Sigma}_k^+(\varepsilon) - \widehat{\Sigma}_k^+(\varepsilon) P_\ell \big(P_\ell' \widehat{\Sigma}_k^+(\varepsilon) P_\ell\big)^+ P_\ell' \widehat{\Sigma}_k^+(\varepsilon)
\end{equation}
for $k, \ell = 1, 2$, and $k \neq \ell$. Then,
\begin{equation}\label{eq:projTest.Est}
    \Gamma_2^\#(\varepsilon) \xrightarrow{\mathcal{D}} \xi, \hspace{0.15cm} \text{ with } \hspace{0.15cm} \xi \sim \chi^2\big(\widehat{r}_2(\varepsilon)\big),
\end{equation}
where $\widehat{r}_2(\varepsilon) = \widehat{r}_{1,2}(\varepsilon) + \widehat{r}_{2,1}(\varepsilon)$ and $\widehat{r}_{k,\ell}(\varepsilon) = \rank(\widehat{\Sigma}_k; \varepsilon) - \rank\big( P_{\ell}' \widehat{\Sigma}_k^+(\varepsilon) P_{\ell} \big)$ for $k, \ell = 1,2$, and $k \neq \ell$. Furthermore, note that $\widehat{r}_2^l(\varepsilon) \leq \widehat{r}_2(\varepsilon) \leq \widehat{r}_2^u(\varepsilon)$ with
$$
\widehat{r}_2^l(\varepsilon) := \rank(\widehat{\Sigma}_1; \varepsilon) + \rank(\widehat{\Sigma}_2; \varepsilon) - 2d, \hspace{0.25cm} \widehat{r}_2^u(\varepsilon) := \rank(\widehat{\Sigma}_1; \varepsilon) + \rank(\widehat{\Sigma}_2; \varepsilon).
$$
Then, with $\xi^l \sim \chi^2(\widehat{r}^l_2(\varepsilon)), ~ \xi^u \sim \chi^2(\widehat{r}^u_2(\varepsilon))$, the p-value based on \eqref{eq:projTest.Est} can be bounded by
\begin{equation}\label{eq:upperBound}
    \mathbb{P} \big( \xi^l > \Gamma_2^\#(\varepsilon) ~|~ H_0\big) \leq \mathbb{P} \big( \xi > \Gamma_2^\#(\varepsilon) ~|~ H_0\big) \leq \mathbb{P} \big( \xi^u > \Gamma_2^\#(\varepsilon) ~|~ H_0\big).
\end{equation}
\end{prop}

We include \eqref{eq:upperBound} to deal with the potentially inconsistent rank estimators of $P_{\ell}' \widehat{\Sigma}_k^+(\varepsilon) P_{\ell}$, $k, \ell = 1,2$ with $k \neq \ell$, and state the following proposition to justify the effectiveness of the relaxed test based on \eqref{eq:upperBound}. In particular, the proposition indicates that the hypothesis gets rejected with high probability within the hypothesis space $H_{1}$.

\begin{prop}\label{prop:alt.behave}
Under the alternative hypothesis $H_1$ in \eqref{hyp:part}, set
\begin{align*}
	& \bm{m}_1 = \Vector (M_1) - P_2 \big(P_2' \Sigma_1^+(\varepsilon) P_2\big)^+ P_2' \Sigma_1^+(\varepsilon) \Vector(M_1),\\
	& \bm{m}_2 = \Vector (M_2) - P_1\big(P_1' \Sigma_2^+(\varepsilon) P_1\big)^+ P_1' \Sigma_2^+(\varepsilon) \Vector(M_2),
\end{align*}
with $\varepsilon$ chosen by \autoref{thm:LLRest} and $\bm{m}_i \in \mathbb{R}^{d^2}$ for $i = 1,2$. If $\widehat{\Sigma}_i^+(\varepsilon) \bm{m}_i \neq 0$ for $i = 1,2$, then the test statistic \eqref{eqn:gamma1.Est} satisfies
$$
\lim_{n \to \infty} \Gamma_2^\#(\varepsilon) \to \infty \hspace{0.15cm} \Rightarrow \hspace{0.15cm} \lim_{n \to \infty} \mathbb{P} \big( \xi > \Gamma_2^\#(\varepsilon) ~|~ H_1\big) = 0.
$$
\end{prop}

Note that under the null hypothesis $H_0$, it might also be true that $\bm{m}_i \neq 0$ when $\widehat{\Sigma}_i^+(\varepsilon)$ is singular, but $\widehat{\Sigma}_i^+(\varepsilon) \bm{m}_i = 0$ for $i = 1,2$ always hold.

\subsubsection{Error analysis}
\label{sec:error}
In this section, we study the effects of replacing the matrices $P_1$ and $P_2$ in \eqref{eqn:Ahat} by their estimators in our proposed test. In particular, we start from the expression \eqref{eqn:polyP} of polynomial basis. We define the estimators \eqref{eqn:polyP} for $P_1$ and $P_2$ as
$$
\widehat{P}_i = (\Vector(A_i^1)/\|A_i^1\|_F, \dots, \Vector(A_i^{d})/\|A_i^d\|_F)
$$
for $i = 1,2$. Then, under \autoref{ass:normal}, $\widehat{P}_1$ and $\widehat{P}_2$ are consistent estimators for $P_1$ and $P_2$ with the same convergence rate $1/c(n)$. However, even with extra care about the covariance singularity, replacing $P_i$ by $\widehat{P}_i$, for $i = 1, 2$, in $\Gamma_2^\#(\varepsilon)$ in \eqref{eqn:gamma1.Est} makes the asymptotic distribution of the test statistic \eqref{eq:projTest.Est} inaccurate. For this reason, one thing remains to be discussed is whether the statistical order of the error introduced from this approximation step is negligible in testing. To be more precise, for \autoref{thm:LLRest}, the error for the first summand in $\Gamma_2^\#(\varepsilon)$ is
\begin{equation} \label{eqn:deltaeps}
    \Delta_\varepsilon := 
    c^2(n) \Vector (A_1)' (\widehat{Q}_{1,2}[\varepsilon] - \widehat{\mathcal{Q}}_{1,2}[\varepsilon]) \Vector(A_1) 
\end{equation}
with $\widehat{Q}_{1,2}[\varepsilon]$ as in \eqref{eq:Qkl} and $\widehat{\mathcal{Q}}_{1,2}[\varepsilon]$ is defined by replacing the matrices $P_{\ell}$ in \eqref{eq:Qkl} by their sample counterparts $\widehat{P}_{\ell}$ such that
\begin{equation} \label{eqn:mathcalQ}
\widehat{\mathcal{Q}}_{k,\ell}[\varepsilon] := \widehat{\Sigma}_k^+(\varepsilon) - \widehat{\Sigma}_k^+(\varepsilon) \widehat{P}_\ell \big(\widehat{P}_\ell' \widehat{\Sigma}_k^+(\varepsilon) \widehat{P}_\ell\big)^+ \widehat{P}_\ell' \widehat{\Sigma}_k^+(\varepsilon)
\end{equation}
for $k, \ell = 1, 2$, $k \neq \ell$. The following proposition provides information about the asymptotic behavior of $\Delta_\varepsilon$ in \eqref{eqn:deltaeps}. The proof can be found in \autoref{app:profs}.

\begin{prop}\label{lem:asyerror} 
Assume the choice of $\varepsilon$ satisfies $\Sigma_1(\varepsilon) = \Sigma_1$. Then, under \autoref{ass:normal}, there exists an $r_{1,2} \times r_{1,2}$ positive semi-definite matrix
$
\widecheck{\Sigma} = \widecheck{\Sigma}(M_1, M_2, \Sigma_1, \Sigma_2)
$
such that the error term in \eqref{eqn:deltaeps} satisfies
$$
\Delta_\varepsilon \xrightarrow{\mathcal{D}} 
Z, \hspace{0.12cm} \text{ with } \hspace{0.12cm} Z= 
\sum_{i = 1}^{r_{1,2}} 2\sqrt{\lambda_i} (\nu_{i,1} - \nu_{i,2}),
$$
where $\lambda_i$ are the eigenvalues of $\widecheck{\Sigma}$, and $\nu_{i,j} \stackrel{i.i.d.}{\sim} \chi^2(1)$ for $i \in  \{1, \dots, r_{1,2}\}, ~ j=1, 2$. Furthermore, the variance of the limit is $\var(Z) = 16 \tr(\widecheck{\Sigma})$.
\end{prop}

According to \autoref{lem:asyerror}, the error term $\Delta_\varepsilon$ in \eqref{eqn:delta1} is still asymptotically unbiased. However, with a mild choice of matrix dimension $d$, its asymptotic variance, which represents the perturbation range, is comparable with the magnitude of the test statistic $\Gamma_2^\#(\varepsilon)$ in \eqref{eq:projTest.Est}, as the matrix $\widecheck{\Sigma} \in \mathbb{R}^{r_{1,2} \times r_{1,2}}$ is generated by well-conditioned matrices $(M_1, M_2, \Sigma_1, \Sigma_2)$. Hence, even with the relaxed test introduced in \autoref{thm:LLRest}, there are no guarantees that the test statistic is valid in real applications. The weighted projections $\widehat{A}_i$, however, could sometimes be useful while problem setup or interests change.

On the other hand, due to the lack of stochastic convergence results for optimization with respect to the common eigenvectors $V$, the consistency rate of plugging $\widehat{V}$ from `\textit{(W)JDTE}' along with its inverse $\widehat{U} = \widehat{V}^{-1}$ into \eqref{eqn:eigvP} for $\widehat{P}_i$ remains unclear. However, as long as the optimization procedure fails to improve the original $1/c(n)$ rate in \Cref{ass:normal} with positive probability, the analogous derivations will lead to a similar conclusion as \autoref{lem:asyerror}.


On the contrary, if one has confident prior knowledge of common eigen-structures, one can simply define the space matrices $P_1$ and $P_2$ using such prior information to make this particular approach applicable with reasonably strong test power. In addition to the direct access to the common eigenvectors for constructing \eqref{eqn:eigvP}, knowledge of common eigen-structures could also be that, when defining \eqref{eqn:polyP}, there is a reference square matrix which shares eigenvectors with the matrices to be tested.

\subsection{Summary of two-sample test}\label{sec:sum}
The test methods developed in \autoref{sec:commute} and \autoref{sec:llrt} could be applied in different settings. For example, if the estimators are available with reasonable asymptotic normality, only the commutator-based test design would guarantee acceptable effectiveness; and if exact eigen-information is given with certainty, the LLR test could be a good choice. However, cases with such strong restrictions and adequate information could be rather rare in real applications. Such statement could be justified in \autoref{sec:simu} by simulations. In the applications later, the commutator-based test is conducted.

\section{Multi-sample test}\label{sec:multi}
An extension of the topic introduced in \autoref{sec:commute} is to conduct the test on a larger pool of matrices ($p \ge 2$). 
In addition to testing simultaneously over all pairs of samples using the methods based on the two-sample test introduced in \autoref{sec:commute}, we are more interested in whether the same hypothesis holds across the whole pool of matrices. With that in mind, we propose to use the estimated optimal common eigenvectors and test whether they annihilate the off-diagonal elements of the matrices after transformation.

For the simultaneous test, with the commutator-based test developed in \autoref{sec:commute}, the approach and the test statistic are straightforward. For instance, a matrix of test statistics (or p-values) represents the pairwise test results and conclusions can then be drawn. Hence, we will omit further details here and focus on the more comprehensive approach regarding whether the same hypothesis holds across the whole pool of matrices. Specifically, we will refer to optimization algorithms for calculating the common eigenvectors that almost diagonalize a pool of matrices (see \autoref{subsec:common}) and then design a test statistic (see \autoref{subsec:eigtest}).

\subsection{Common eigenvectors finder}\label{subsec:common}
In this section, we briefly introduce the setup of the optimization problem of finding the optimal diagonalizer for a pool of matrices.

Recall our setup, with a pool of matrix-valued statistics $\mathcal{A} = \{A_i\}_{i = 1}^p$ estimating the matrices $\mathcal{M} = \{M_i\}_{i = 1}^p$. We refer to the '\textit{(W)JDTE}' algorithm by \cite{andre} due to its performance in terms of speed and accepted accuracy. The algorithm provides the common eigenvectors of a pool of matrices by minimizing the objective function
\begin{equation}
    \text{off}(U; \mathcal{A}) = \sum_{i = 1}^p \text{off}_2(U^{-1} A_i U).
\end{equation}
Here, $\text{off}_2(X) = \sum_{i \ne j} |X_{i,j}|^2$ denotes the off-diagonal sum-of-squares for $X \in \mathbb{R}^{d \times d}$. For the details of the algorithm; see \cite{andre}.

\subsection{Eigenvector test}\label{subsec:eigtest}
In this section, we propose a test for $H_0$ in \eqref{hyp:all} allowing the number of matrices $p$ to be larger than two. The test is based on the assumption that an invertible matrix $V \in \R^{d \times d}$ is given as a guess for the common eigenvector matrix. 
We acknowledge that the assumption of knowing the matrix $V$ is quite restrictive. For all practical purposes, we refer to the optimization problem in \autoref{subsec:common} which provides an optimal diagonalizer based on the matrix estimators $\mathcal{A} = \{A_i\}_{i = 1}^p$.
From a theoretical perspective, existing literature does not suggest any formal testing procedure based on explicit estimators for $V$. In particular, there are no asymptotic results, neither under the assumption that the matrices to be tested are asymmetric nor symmetric. For this reason, we see the below stated theoretical results under the assumption that $V$ is known as a starting point and leave a more rigorous investigation for future work.


Define the function
$
\offvec_d: \R^{d \times d} \to \R^{d(d - 1)}, ~ \offvec_d(X) = S_d \Vector(X),
$
which stacks all off-diagonal elements of a square matrix $X$ with dimension $d$ columnwise. We will always use $S_d$ as the off-diagonal selection matrix for square matrices of dimension $d$. The test can then be designed as follows.

\begin{prop}\label{thm:multi_eig}
Suppose \Cref{ass:normal,ass:distinct}, and $V$ is given as a guess for the common eigenvector matrix. Let 
\begin{equation} \label{eq:xii}
\bm{\zeta}_i = \offvec_d(V^{-1} A_i V), \hspace{0.2cm} i = 1, \dots, p, 
\hspace{0.2cm}\text{and}\hspace{0.2cm} 
S_{V,d} = S_d (V' \otimes V^{-1}). 
\end{equation}
Then, under $H_0$ in \eqref{hyp:all},
$
    c(n) \bm{\zeta}_i \xrightarrow{\mathcal{D}} \mathcal{N}(0, \Theta_i)
$
with $\Theta_i = S_{V,d} \Sigma_i S_{V,d}'$ for $i = 1, \dots, p$.
The test statistic is defined as and satisfies
\begin{equation}\label{eqn:multi_wald}
    \Gamma_3 := c^2(n) \sum_{i = 1}^p \bm{\zeta}_i' \Theta_i^+ \bm{\zeta}_i \xrightarrow{\mathcal{D}} \chi^2(r_3),
\end{equation}
where $r_3 = \sum_{i=1}^p \rank(\Theta_i)$.
\end{prop}

Note that in practice, the eigenvector matrix to be tested is always obtained from optimization, and hence this idea highly depends on the accuracy of such algorithms.
In order to reduce the influence of estimation errors, we develop the following analogous test that tolerates relatively larger errors while maintaining acceptable efficiency.
\begin{cor}\label{cor:multi}
Under the assumptions of \autoref{thm:multi_eig}, let $\Theta = \blkdiag(\{\Theta_i\}_{i = 1}^p)$ be a block-diagonal matrix, and
$
\bm{\zeta} := \big(\bm{\zeta}_1', \dots, \bm{\zeta}_p' \big)' \in \mathbb{R}^{p(d^2-d)},
$ with $\bm{\zeta}_{i}$ as in \eqref{eq:xii} such that $c(n) \bm{\zeta} \xrightarrow{\mathcal{D}} \mathcal{N}(0, \Theta)$. Then, the test statistic is defined as and satisfies
\begin{equation}\label{eqn:multi_norm}
    \Gamma_3^* := c^2(n) \sum_{i=1}^p \|\bm{\zeta}_i\|^2
    \xrightarrow{\mathcal{D}} \psi_3^* \coloneqq
    \sum_{r=1}^{p(d^2-d)} \lambda_{r}(\Theta) \chi^2(1),
\end{equation}
where $\lambda_{r}(\Theta)$ denotes the $r$th eigenvalue of $\Theta$. Furthermore, the p-value based on \eqref{eqn:multi_norm} can be approximated by
\begin{equation}\label{eqn:full_approx}
    \mathbb{P}(\psi_3^* > \Gamma_3^* ~|~ H_0) \approx \mathbb{P}(\gamma_3^* > \Gamma_3^* ~|~ H_0), \hspace{0.2cm} \mbox{where} \hspace{0.2cm} \gamma_3^* \sim Gamma\Big(\frac{\tr(\Theta)^2}{2 \tr(\Theta^2)}, \frac{\tr(\Theta)}{2 \tr(\Theta^2)}\Big).
\end{equation}
\end{cor}
The approximation in \eqref{eqn:full_approx} between $\psi_3^*$, a weighted sum of chi-squared distributed random variables, and $\gamma_3^*$, a gamma distribution, was introduced in \citet[Theorem 3.1]{box1954} based on matching first- and second-order moments of the two distributions. For accuracy analysis, see \cite{bodenham2016comparison}.

Note, that \autoref{thm:multi_eig} requires to calculate the generalized inverse of $\Theta$.
In contrast, \autoref{cor:multi} only needs the trace of $\Theta$. Calculating the trace of $\Theta$ is expected to be more robust towards estimation errors of the eigenvector matrix $V$ than finding the generalized inverse of $\Theta$.

In practice, only the estimated $\widehat{\Sigma}_i$ are accessible. Due to possible singularity issues with the general inverse $\Theta_i^+$ in \autoref{thm:multi_eig}, we use results in \autoref{app:est} to introduce the following proposition.

\begin{prop}\label{thm:multi_eig.Est}
Suppose \Cref{ass:normal,,ass:covConsistent,,ass:distinct}. Then, for a given threshold $\varepsilon > 0$ that is not an eigenvalue of $\Theta_i$ for $i = 1, \dots, p$, 
the test statistic is defined as and satisfies
\begin{equation}\label{eq:gamma3.est}
    \Gamma_3^\#(\varepsilon) := c^2(n) \sum_{i = 1}^p \bm{\zeta}_i' \widehat{\Theta}_i^+(\varepsilon) \bm{\zeta}_i \xrightarrow{\mathcal{D}} \chi^2(\widehat{r}_3(\varepsilon)),
\end{equation}
where $\widehat{\Theta}_i = S_{V, d} \widehat{\Sigma}_i S_{V, d}$ and $\widehat{r}_3(\varepsilon) = \sum_{i=1}^p \rank(\widehat{\Theta}_i; \varepsilon)$.
\end{prop}

\section{Partial test}\label{sec:part}
In this section, we focus on the hypothesis $H_0^{*}$ in \eqref{hyp:part}. We first reformulate the hypothesis testing problem (\autoref{sec:partialprob}) and then design the corresponding test statistics and analyze their asymptotic behavior (\autoref{sec:partialtest}). Meanwhile, an optimization algorithm is also proposed to approximate the matrix $V$ in the statement of $H_0^{*}$ (\autoref{sec:partalg}).

\subsection{Problem representation} \label{sec:partialprob}
Suppose $\{M_i\}_{i = 1}^p$ satisfy $H_0^{*}$, then for $V$ specified in the hypothesis, there exists an orthogonal $d \times k$ matrix $Q_k$ (i.e. $Q_k' Q_k = I_k$), such that $V = Q_k R$, where $R$ is a $k \times k$ upper-triangular matrix. By orthogonally spanning $Q_k = (\bm{q}_1, \dots, \bm{q}_k)$ to $Q = (\bm{q}_1, \dots, \bm{q}_d) \in \mathbb{R}^{d \times d}$, we have
\begin{equation}
   Q' M_i Q = \begin{pmatrix} R D_i R^{-1} & * \\ 0 & ** \end{pmatrix} ~ =: ~ \begin{pmatrix} \Phi_i & * \\ 0 & ** \end{pmatrix}\label{eqn:part}
\end{equation}
with $D_{i}$, $i=1,\dots,p$, as in \eqref{hyp:part}.
We use the symbols $*$ and $**$ to denote non-zero block matrices with proper dimensions which are not relevant. Note that the set of upper-triangular matrices $\{\Phi_i\}_{i = 1}^p$ shares the common eigenvectors $R$. For the estimators $\mathcal{A} = \{A_i\}_{i = 1}^p$ given by \autoref{ass:normal}, suppose a matrix $\widehat{Q}$ is given as a guess for $Q$, such that
\begin{equation}
    \widehat{Q}' A_i \widehat{Q} = \begin{pmatrix} B_i & * \\ C_i & ** \end{pmatrix},\label{eqn:part_noisy}
\end{equation}
where $B_i \in \mathbb{R}^{k \times k}$ and $C_i \in \mathbb{R}^{(d-k) \times k}$. 
The matrix $\widehat{Q}$ can either be given by knowing the ground-truth $Q$ or by estimation; see \autoref{sec:partalg} for an algorithm to estimate $Q$.
The procedure of our test starts from finding the orthogonal matrix $Q$ that contains the first $k$ columns as the common invariant subspace, and designing tests on the transformed matrices $\{B_i, C_i\}_{i = 1}^p$. The test is a combination of:
\begin{enumerate}
    \item Whether the transformation by the orthogonal matrix $\widehat{Q}$ introduces the all-zero lower-left block in \eqref{eqn:part} by testing on $\{C_i\}_{i = 1}^p$.
    
    \item Whether the $k \times k$ matrices $\{\Phi_i\}_{i = 1}^p$ defined in \eqref{eqn:part} satisfy $H_0$ in \eqref{hyp:all} by testing on $\{B_i\}_{i = 1}^p$.
\end{enumerate}
To introduce our test statistic, suppose $\widehat{Q}$ is given, and let
\begin{equation*}
\widehat{Q}_{BC} = \begin{pmatrix} S_B \\ S_C \end{pmatrix} (\widehat{Q}' \otimes \widehat{Q}')
\hspace{0.2cm}
\text{ with }
\hspace{0.5cm}
\begin{matrix}
S_B \mbox{vec}(\widehat{Q}' A_i \widehat{Q}) = \mbox{vec}(B_i), \\
S_C \mbox{vec}(\widehat{Q}' A_i \widehat{Q}) = \mbox{vec}(C_i),
\end{matrix}
\end{equation*}
where $S_B$ and $S_C$ are selection matrices defined 
according to \eqref{eqn:part_noisy}. Hence, the partially common eigenvectors of $\{M_i\}_{i = 1}^p$ can be estimated as $\widehat{V} = \widehat{Q}_k \widetilde{V}$, where $\widehat{Q}_k$ is a matrix of the first $k$ columns of $\widehat{Q}$ and $\widetilde{V}$ the estimated common eigenvectors of $\{B_i\}_{i = 1}^p$. Referring to \autoref{subsec:common}, $\widetilde{V}$ can be received from '\textit{(W)JDTE}'. So far, we have supposed that $\widehat{Q}_k$ is a guess. However, it can be estimated by an algorithm proposed in \autoref{sec:partalg} below.


\subsection{Partial eigenvector test}\label{sec:partialtest}
In this section, we introduce our test statistic to test for partially common eigenvectors and present their asymptotic behavior.

\begin{prop}\label{thm:part}
Suppose \Cref{ass:normal,ass:distinct} are satisfied and let
\begin{equation}\label{eqn:wii}
\bm{w}_i =  P_w \Vector(A_i) 
\hspace{0.2cm} \text{ with } \hspace{0.2cm}
P_w := \begin{pmatrix} S_{\widetilde{V}, k} & 0 \\ 0 & I_{k(d-k)} \end{pmatrix} ~ Q_{BC},
\end{equation}
where $S_{\widetilde{V}, k}$ is defined as in \eqref{eq:xii} and $Q_{BC}$ is analogous to $\widehat{Q}_{BC}$ except replacing the guess $\widehat{Q}$ with true $Q$.
Then, under $H_{0}$ in \eqref{hyp:all},
$
    c(n) \bm{w}_i \xrightarrow{\mathcal{D}} \mathcal{N}(0, \Omega_i) \label{eqn:partnorm}
$
with $\Omega_i = P_w \Sigma_i P_w'$ for $i = 1, \dots, p$. The test statistic is defined as and satisfies
\begin{equation}
    \Gamma_4 := c^2(n) \sum_{i = 1}^p \bm{w}_i' \Omega_i^+ \bm{w}_i \xrightarrow{\mathcal{D}} \chi^2(r_4),\label{eqn:partchi}
\end{equation}
where $r_4 = \sum_{i=1}^p \rank(\Omega_i)$.
\end{prop}

And similar to \autoref{cor:multi}, we design the following test statistic with p-value approximated by a gamma distribution.
\begin{cor}\label{cor:partial}
Under the assumptions in \autoref{thm:part}, let $\Omega = \blkdiag(\{\Omega_i\}_{i = 1}^p)$ be a block-diagonal matrix, and
$
\bm{w} := \big(\bm{w}_1', \dots, \bm{w}_p' \big)' \in \mathbb{R}^{pk(d-1)},
$ with $\bm{w}_{i}$ as in \eqref{eqn:wii} such that $c(n) \bm{w} \xrightarrow{\mathcal{D}} \mathcal{N}(0, \Omega)$. Then the test statistic is defined as and satisfies
\begin{equation}\label{eqn:partgam}
    \Gamma^*_4 := c^2(n) \|\bm{w}\|^2
    \xrightarrow{\mathcal{D}} \psi_4^* \coloneqq
    \sum_{r=1}^{pk(d-1)} \lambda_{r}(\Omega) \chi^2(1),
\end{equation}
where $\lambda_r(\Omega)$ denotes the $r$th eigenvalue of $\Omega$. Furthermore, the p-value based on \eqref{eqn:partgam} can be approximated by
\begin{equation}\label{eqn:part_approx}
    \mathbb{P}(\psi_4^* > \Gamma_4^* ~|~ H_0) \approx \mathbb{P}(\gamma_4^* > \Gamma_4^* ~|~ H_0), \hspace{0.2cm} \mbox{where} \hspace{0.2cm} \gamma_4^* \sim Gamma\Big(\frac{\tr(\Omega)^2}{2 \tr(\Omega^2)}, \frac{\tr(\Omega)}{2 \tr(\Omega^2)}\Big).
\end{equation}
\end{cor}

Similar as \autoref{thm:multi_eig.Est}, we use results in \autoref{app:est} for the following tractable version of \autoref{thm:part} that takes care of possible singularity issues with $\Omega_i^+$ in \eqref{eqn:partchi}.

\begin{prop}\label{thm:part.Est}
Assume \Cref{ass:normal,,ass:covConsistent,,ass:distinct} are satisfied. Then, for a given threshold $\varepsilon > 0$ that is not an eigenvalue of $\Omega_i$ for $i = 1, \dots, p$, 
the test statistic is defined as and satisfies
\begin{equation}\label{eq:gamma4.est}
    \Gamma_4^\#(\varepsilon) := c^2(n) \sum_{i = 1}^p \bm{w}_i' \widehat{\Omega}_i^+(\varepsilon) \bm{w}_i \xrightarrow{\mathcal{D}} \chi^2(\widehat{r}_4(\varepsilon)),
\end{equation}
where $\widehat{\Omega}_i = P_w \widehat{\Sigma}_i P_w'$ for $i = 1, \dots, p$, and $\widehat{r}_4(\varepsilon) = \sum_{i=1}^p \rank(\widehat{\Omega}_i; \varepsilon)$.
\end{prop}

\subsection{Optimization algorithm} \label{sec:partalg}
This section is dedicated to finding an estimator $\widehat{Q}$ for $Q$ in \eqref{eqn:part}.
In implementation, $\widehat{Q}$ can be obtained by minimizing the following objective function
$$
f(Q; \mathcal{A}, k) = \sum_{i = 1}^p \sum_{r = k+1}^d \sum_{s = 1}^k (\bm{q}_r' A_i \bm{q}_s)^2 ~ \mbox{ subject to } \bm{q}_r' \bm{q}_s = \delta_{rs}, ~ \forall~ r,s = 1, \dots, d,
$$
where $Q = (\bm{q}_1, \dots, \bm{q}_d) \in \mathbb{R}^{d \times d}$.
\cite{tensor} introduced a version of Gauss-Newton algorithm for the joint Schur decomposition based on matrix exponential, and showed its global minimum guarantees if the initial value is sufficiently close to the ground-truth $Q$. In our work, we inherit the idea of this algorithm with slight revision, where the major difference is to substitute the selection matrix from lower-triangular indicator to the $(d-k) \times k$ lower-left block indicator.

We also introduce a warm-up algorithm to supply the initial values for this Gauss-Newton approach. Since the matrices $\{\Phi_i\}_{i = 1}^p$ in \eqref{eqn:part} are upper-triangular, we can split the minimization with respect to $Q$ into sequentially optimizing each column $\bm{q}_r$ with $r$ from $1$ to $k$ based on the following objective function
$$
f^*(Q; \mathcal{A}, k) = \sum_{i = 1}^p \sum_{s = 1}^k \sum_{r = s+1}^d (\bm{q}_r' A_i \bm{q}_s)^2, \mbox{ with } Q = (\bm{q}_1, \dots, \bm{q}_d) \in \mathbb{R}^{d \times d}.
$$
More precisely, we introduce the following \autoref{algo:Qhat} as the whole process for optimizing orthogonal $\widehat{Q}$, including the initialization warm-up before \autoref{step:GN}.

\begin{algorithm}[htbp]
    \caption{Estimation of $Q$}\label{algo:Qhat}
    \KwIn{Estimators $\mathcal{A} = \{A_i\}_{i=1}^p$, number of common eigenvectors $k$.}
    \KwOut{Estimator $\widehat{Q}$ for $Q$.}
    Initialize $\mathcal{O}_0 = I_d $ as an identity matrix.\\
    \For{$j = 1:k$}{
        Optimize $\mathcal{O}_j = \argmin_Q f(Q; \mathcal{A}, 1)$ and write $\mathcal{O}_j = (\bm{p}_j, P_j)$, where $\bm{p}_j \in \mathbb{R}^{d-j+1}$ and $P_j \in \mathbb{R}^{(d-j+1) \times (d-j)}$. \label{algo:opt}\\
        Update $\mathcal{O}_{j-1} = \mathcal{O}_{j-1} \mathcal{O}_j$.\\
        Update $\mathcal{A}$ by $A_i = P_j' A_i P_j$ for $i = 1, \dots, p$.
    }
    Input $\mathcal{O}_k$ into the modified version of Gauss-Newton approach by \cite{tensor} as the initial value and get the final output $\widehat{Q}$.\label{step:GN}
\end{algorithm}
For realization of the optimization on \autoref{algo:opt}, we refer to the FG-algorithm by \cite{Flury86}.

\section{Extension to symmetric matrices}
\label{sec:ext}

As mentioned in the introduction, the analysis of common eigenvectors has many applications for symmetric matrices, for example, CPCA. The test methods introduced in this work can be implemented for symmetric matrices as well if we take additional care of the assumptions in \autoref{subsec:ass}.

Suppose the matrices $M_i$, $i=1,\dots,p$, and their respective estimators $A_i$, $i=1,\dots,p$, are symmetric matrices. Denote the function $\Vech: \mathbb{R}^{d \times d} \to \mathbb{R}^{d(d+1)/2}$ that converts a symmetric matrix $A$ into a vector stacking only distinct elements columnwise. Then, from estimations for symmetric $M_i$, the available consistency statements are of the form
$$
c(n) \Vech(A_i - M_i) \xrightarrow{\mathcal{D}} \mathcal{N}(0, \widetilde{\Sigma}_i)
$$
with positive semi-definite $\widetilde{\Sigma}_i \in \mathbb{R}^{d(d+1)/2 \times d(d+1)/2}$. There exists the duplication matrix $G_d \in \{0, 1\}^{d^2 \times d(d+1)/2}$ such that $G_d \Vech(A) = \Vector(A)$ for any symmetric $A \in \mathbb{R}^{d \times d}$; see \cite{magnus2019matrix} for more details on such operations. Hence, we can obtain exactly the same setup as \autoref{ass:normal}, since
$$
c(n) \Vector(A_i - M_i) = c(n) G_d \Vech(A_i - M_i) \xrightarrow{\mathcal{D}} \mathcal{N}(0, \Sigma_i)
$$
with $\Sigma_i = G_d \widetilde{\Sigma}_i G_d'$. It is then straightforward to implement the above test designs directly, except that we may require the input eigenvector matrix $V$ (or $\widehat{V}$) to be orthogonal. Such orthogonal matrices can be obtained referring to existing optimization schemes like FG-algorithm by \cite{Flury86}.

As we focus on the general asymmetric setting, our simulation study as well as the application section do not cover symmetric extensions.

\section{Simulation study}\label{sec:simu}
Simulation studies are run for two-sample, multi-sample, and partial tests. The barplots of p-values from multiple replicates, especially, the type I and type II errors which are readable from the plots of histograms, are given as evidence for the effectiveness of our test designs. Complementary to the plots presented in this section, we refer the reader to \autoref{se:compl1} in the supplementary material for tables providing type I and II errors for all our simulation studies.

In addition, instead of implementing \Cref{thm:comm,,thm:LLR,,thm:multi_eig,,thm:part}, we always turn to their truncated versions (\Cref{thm:comm.Est,,thm:LLRest,,thm:multi_eig.Est,,thm:part.Est} respectively) in \Cref{sec:simu,,sec:app}, in order to take care of possibly singular covariance matrices. Furthermore, since the following examples with sample size $n$ all have the same rate of convergence $\sigma(n) = n^{-1/2}$ for their limiting covariance matrix estimators, we will not further specify but choose $\varepsilon = n^{-1/3}$ by default whenever applicable in implementations. This choice of $\varepsilon$ is also in accordance with the discussion in \autoref{rem:epsilon.order} and maintains the test power.

\subsection{Data generating process (DGP)}\label{subse:DGP}
Prior to our testing analysis, we first introduce the data generating process (DGP) for our targeted matrices $\mathcal{M}_p(\rho, k; d) = \{M_i(\rho, k; d)\}_{i=1}^p$ with parameters $\rho$ accounting for the signal-to-noise variance ratio ($\SNR \coloneqq \frac{1}{\rho^2}$) and $k$ the number of common eigenvectors.

\begin{algorithm}[htbp]
    \KwIn{Number of matrices $p \in \mathbb{N}$, matrix dimension $d \in \mathbb{N}$, number of common eigenvectors $k \in \{1, \dots, d\}$, noise measure $\rho \ge 0$.}
    \KwOut{Pool of matrices $\mathcal{M}_p(\rho, k; d)$ with common eigenvectors.}
    Randomly generate an eigenvector matrix $V(k; d) \in \R^{d \times k}$ ($k \le d$).\label{step:commonV}\\
    Set $V_i(k; d) = V(k; d)$ if $k = d$, otherwise span $V_i(k; d) = (V(k; d), \widetilde{V}_i) \in \R^{d \times d}$ with random but sufficiently distinct $\widetilde{V}_i \in \R^{d \times (d-k)}$ for $i = 1, \dots, p$.\\
    \For{$i=1, \dots, p$}{perturb the $i$-th eigenvector matrix $V_i(k; d)$ as $V_i(\rho, k; d) = V_i(k; d) + \rho E_i$ with noise $E_i$ to be independent and standard normal, element-wise.\label{step:diffV}}
    Randomly generate non-singular diagonal matrices $D_i \in \R^{d \times d}$ for $i = 1, \dots, p$.\\
    Generate the target matrices $\mathcal{M}_p(\rho, k; d) = \{ V_i(\rho, k; d) D_i V_i^{-1}(\rho, k; d)\}_{i=1}^p$.
    \caption{DGP for $\mathcal{M}_p(\rho, k; d)$}\label{algo:DGP}
\end{algorithm}

With the set of target matrices $\mathcal{M}_p(\rho, k; d)$, we can then proceed to design different sampling setup for their corresponding consistent estimators $\mathcal{A}_p(\rho) = \{A_i(\rho)\}_{i=1}^p$. Note that we define $\SNR = \frac{1}{\rho^2}$ to quantify the similarity among the underlying eigenvectors $\{V_i(\rho, k; d)\}_{i=1}^p$ (see \autoref{step:diffV} of \autoref{algo:DGP}) rather than the estimation accuracy for $\mathcal{A}_p(\rho)$, i.e., if ideally $\SNR = \infty$ or $\rho = 0$, the matrices in $\mathcal{M}_p(\rho, k; d)$ can be perfectly diagonalized with common eigenvectors (partially when $k < d$), while the randomness of our test statistics still exists due to the subsequent estimation for $\mathcal{A}_p(\rho)$.

\subsection{Two-sample test} \label{subse:2sample}
With $\mathcal{M}_2(\rho, d; d) = \{M_1(\rho, d; d), M_2(\rho, d; d)\}$ generated following \autoref{algo:DGP} in \autoref{subse:DGP}, we treat them as the mean matrices for two pools of $n$ normally distributed observations and propose the averages of samples as the corresponding estimators $\mathcal{A}_2(\rho)$. The covariance matrices are consistently estimated as well. According to the central limit theorem (CLT), the convergence rate in \autoref{ass:normal} satisfies $c(n) = \sqrt{n}$. If not further specified, we stick to the classical p-value test framework and always reject the null $H_0$ (or $H_0^{*}$) when the p-value is below a $0.05$ significance level.


We set the dimension of the matrices as $d = 5$, $\SNR \in \{10, 1000, \infty\}$, and the sample size $n \in \{50, 250, 1000\}$. For each data-generating setting determined by $(\SNR, n)$, the \autoref{algo:DGP} is repeated $500$ times independently to get the data of $500$ pairs of estimators; see \Cref{fig:comm} for performance of commutator-based test and \autoref{fig:LLR} for LLR test. Note that the first bar in these figures represents the proportion of the 500 p-values lower than the critical threshold 0.05, i.e., the proportion of simulations that one rejects the null hypothesis $H_0$ based on the test design. Ideally for $\SNR = \infty$, this proportion, also known as type I error rate, should be close to 0.05, which is a typical choice of significance level. For $\SNR \ne \infty$, the proportion of p-values outside the first bar, also known as type II error rate, should approach 0 if the test has high power.

\begin{figure}[htbp] 
    \centering
    \centerline{\input{Plots/PvalueCommutator.tikz}}
    \caption{The histograms of p-values for the commutator-based test in \autoref{thm:comm.Est} with simulated data.}
   \label{fig:comm}
\end{figure}

\begin{figure}[htbp] 
    \centering
    \centerline{\input{Plots/PvalueLLR.tikz}}
    \caption{The histograms of p-values for the LLR test. The first / third row are results based on \autoref{thm:LLRest} given the exact polynomial basis \eqref{eqn:polyP} / eigenvector basis \eqref{eqn:eigvP}, while the second / fourth row are obtained by plugging estimated $\mathcal{A}_2(\rho)$ / optimized $\widehat{V}$ into \eqref{eqn:polyP} / \eqref{eqn:eigvP} for implementations of the first / third row.}
   \label{fig:LLR}
\end{figure}

For the commutator-based test \autoref{thm:comm.Est}, we see from \autoref{fig:comm} that with sample size increasing, the p-values of samples from null space ($\SNR = \infty$) tend to be uniformly distributed on the interval $[0,1]$, and the p-values of samples from alternative spaces start to concentrate in the interval $[0, 0.05)$. When the sample size $n$ exceeds a certain level, $200$ for instance, the test performs well with acceptable type I error and excellent type II error.

For the LLR test, we conduct (i) \autoref{thm:LLR} given either the exact $\mathcal{M}_2(\rho,d;d)$ for \eqref{eqn:polyP} or the exact eigenvectors $V$ for \eqref{eqn:eigvP}, and (ii) \autoref{thm:LLRest} with estimators $\mathcal{A}_2(\rho)$ plugged-in. By comparing the first and the third rows of \autoref{fig:LLR} with \autoref{fig:comm}, \autoref{thm:LLR} has better performance in terms of type II error, especially with small sample sizes. With the plugin version \autoref{thm:LLRest} (the second and the fourth rows), however, one frequently fails to distinguish the null hypothesis even under the ideal case when $\SNR = \infty$ or $\rho = 0$. It is compatible with our error analysis given in \autoref{sec:error}. To summarize, though with the well-behaved p-values of \autoref{thm:LLR} under $H_0$, such idea from \autoref{sec:llrt} may rarely be applicable unless the test is against some deterministic reference eigen-structure.

\subsection{Multi-sample test} \label{se:sim-multi}
For the simulation study of the multi-sample test introduced in \autoref{sec:multi}, we follow \autoref{algo:DGP} to set the mean matrices of normally distributed observations as $\mathcal{M}_p(\rho, d; d)$. We consider here the setting that dimension $d = 4$, $p = 8$, $\SNR = \frac{1}{\rho^2} \in \{10, 100, 1000, \infty\}$, and sample size $n \in \{10^2, 10^3, 10^4, 10^5\}$. Each consistent estimate in $\mathcal{A}(\rho)$ is the empirical average from the $n$ observations, hence $c(n) = \sqrt{n}$ by the CLT. Note that when $\rho = 0$, equivalently $\SNR = \infty$, the matrices in $\mathcal{A}_p(\rho)$ satisfy $H_0$, i.e., share common eigenvectors.

\begin{sidewaysfigure}[p] 
    \centering
    \centerline{\input{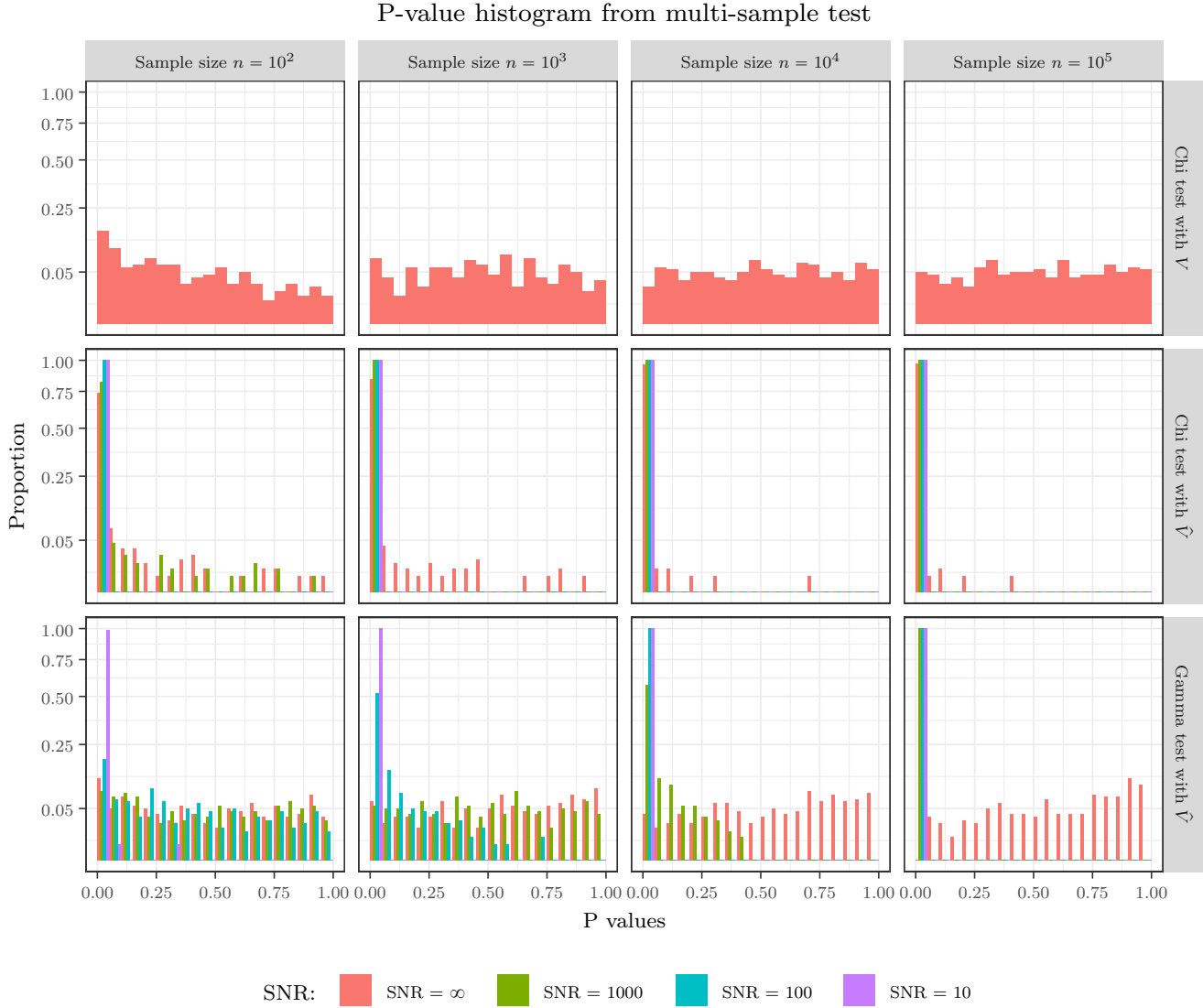}}
    \caption{The histograms of p-values for the multi-sample test. We randomly sampled $200$ independent replicates for each data-generating setting of $(\SNR, n)$. The first row of plots are the test results of \autoref{thm:multi_eig.Est} on $\mathcal{A}_p(0)$ supposing the exact common eigenvectors matrix $V$ is known, hence p-values are only available with $\SNR = \infty$. The simulations of the next two rows use the estimated $\widehat{V}$ from '\textit{(W)JDTE}' as input for \autoref{thm:multi_eig.Est} and \autoref{cor:multi}, respectively.}
   \label{fig:multi}
\end{sidewaysfigure}

To verify the testing efficiency, we (i) use the exact $V = V(d; d)$ (see \cref{step:commonV} of \autoref{algo:DGP}) to test on $\mathcal{A}_p(0)$ according to \autoref{thm:multi_eig.Est}, and (ii) use the optimized $\widehat{V}$ from algorithm '\textit{(W)JDTE}' and implement both \autoref{thm:multi_eig.Est} and \autoref{cor:multi}. We present the test power versus $\SNR$ and sample size $n$ through histogram plots; see \autoref{fig:multi}.

With evidence that the p-values are almost uniformly distributed by using exact $V$, the test of \autoref{thm:multi_eig.Est} is shown to be effective when the supplied common eigenvectors are sufficiently accurate. However, for estimated $V$, one can observe that for small p-values the proportion of rejections exceeds the nominal test size of $0.05$ for the chi-squared test in \autoref{thm:multi_eig.Est}. While the gamma test in \autoref{cor:multi} admits almost the opposite behavior subceeding the nominal test level. From a theoretical perspective that might also be due to the singular sensitivities when inverting the estimated covariance matrices to calculate the test statistic in \autoref{thm:multi_eig.Est}.

As for \autoref{cor:multi}, it is a reasonably efficient test method as the false positive rate (type I error) remains at a relatively low level regardless of the sample size $n$, while the false negative rate (type II error) is following a reasonable pattern that with higher $\SNR$, i.e. less perturbations to the shared eigenvectors, the test is ultimately able to reject the $H_0$ as the sample size $n$ increases to be sufficiently large.

\subsection{Partial test} \label{subse:partialtest}
For the simulations regarding the partial test in \autoref{sec:part}, we fix $k<d$, and generate the mean matrices as $\mathcal{M}_p(\rho, k; d)$ following \autoref{algo:DGP}. Consistent estimates $\mathcal{A}_p(\rho)$ are obtained similarly from the average of the random observations to follow ordinary CLT. We set $d = 4$, $p = 8$, $k = 2$, $\SNR = \frac{1}{\rho^2} \in \{10, 100, 1000, \infty\}$, and sample size $n \in \{10^2, 10^3, 10^4\}$.

\begin{figure}[htb] 
    \centering
    \centerline{\input{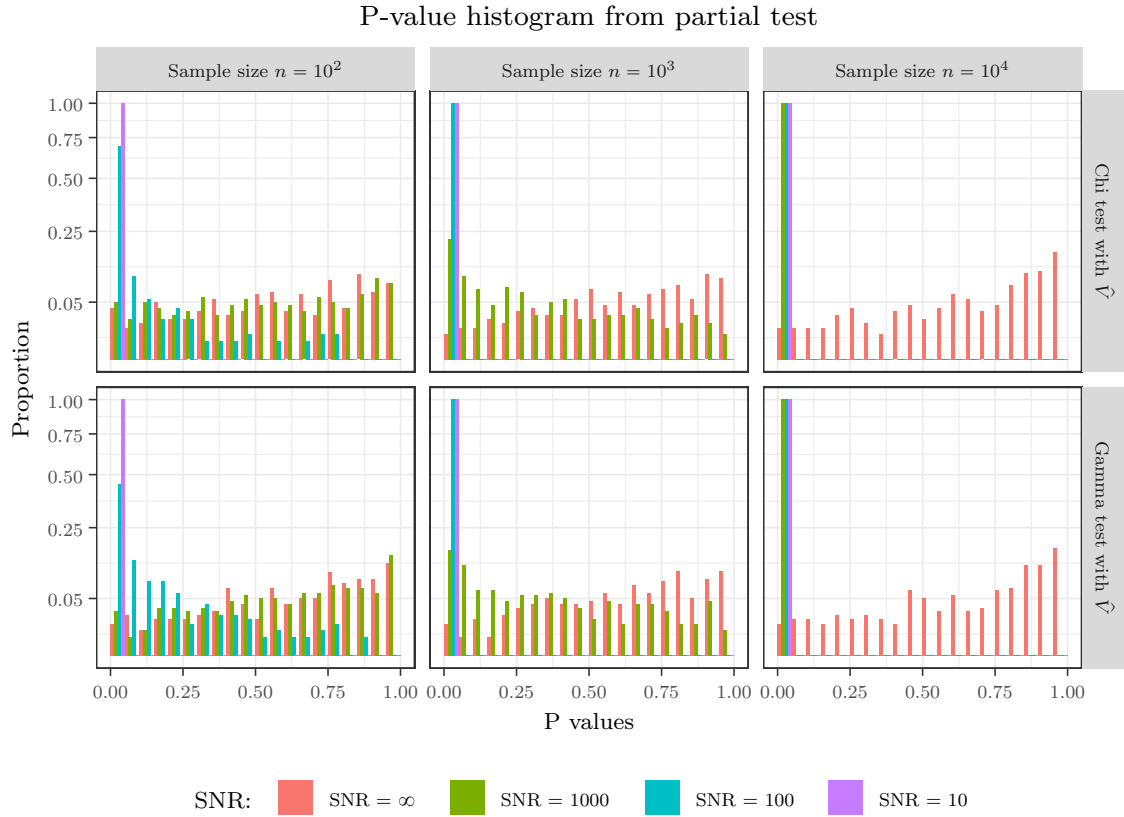}}
    \caption{The histograms of p-values for the partial test. $200$ independent replicates are sampled for each data-generating setting of $(\SNR, n)$. Testing with estimated $\widehat{V}$, the first and second rows are results from \autoref{thm:part.Est} and \autoref{cor:partial}, respectively.}
   \label{fig:partial}
\end{figure}

The histograms in \autoref{fig:partial} show that both \autoref{thm:part.Est} and \autoref{cor:partial} have similar testing power. In addition, even if $\SNR$ increases, i.e. the perturbation on common eigenvectors $V$ becomes subtle, the type II error can maintain at almost zero even with small sample size $n$.

As pointed out in \autoref{sec:part}, we assume that the number of partial common eigenvectors in known. Since this assumption is not feasible in practice, we propose a sequential testing procedure. We refer to \autoref{se:compl2} in the supplementary material for a detailed description of the testing procedure and a corresponding simulation study to access its performance.

\section{Applications} \label{sec:app}
In \autoref{se:VAR} we consider VAR models and analyze their dynamic structure in terms of our test methods. 
In \autoref{se:Markov}, we introduce the application of our partial test on identical stationary distributions of different Markov chains.

\subsection{VAR models} \label{se:VAR}
\begin{figure}[htbp] 
    \centering
    \centerline{\input{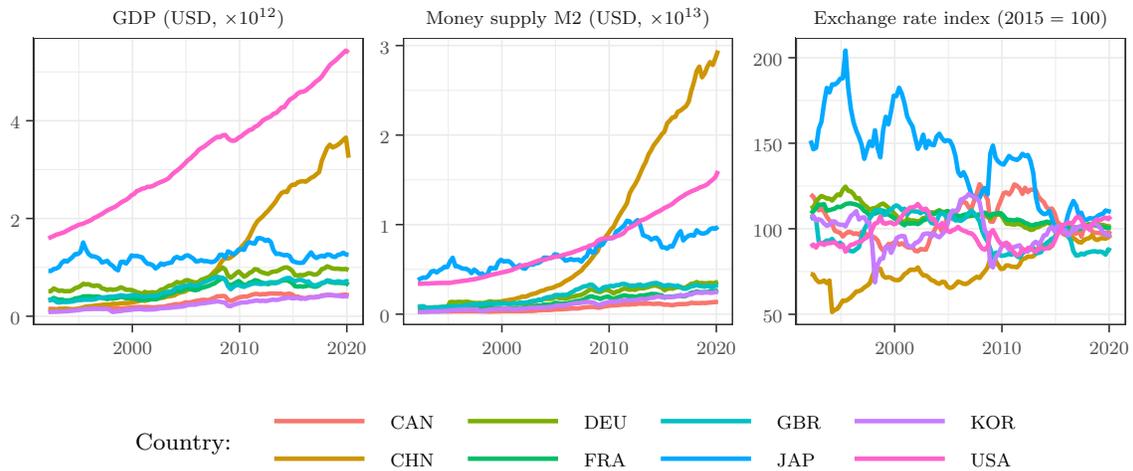}}
    \caption{The time plots of the macro-economic indices of eight countries from the first quarter of 1992 to the first quarter of 2020. The data set is collected from \cite{ceic} and \cite{oecd}.}
    \label{fig:VAR1_eg}
\end{figure}

We consider gross domestic product (GDP), money supply (M2), and real effective exchange rate (REER) for eight of the most influential countries distributed across three different continents; see \Cref{fig:VAR1_eg}. GDP and M2 data are available through \cite{ceic}, and REER data through \cite{oecd}. The Bayesian information criterion (BIC) favors VAR models of order one for all eight countries. Therefore, for each country, we fit 
$$
\bm{y}_{t}^c = \bm{\mu}^c + \Phi^c \bm{y}_{t - 1}^c + \bm{e}_{t}^c,
\hspace{0.2cm} t = 1, \dots, n+1,
$$
to the three different economic indices (GDP, M2, REER),
where the superscript $c$ distinguishes the eight countries.
We may compare the growth tendency among subjects if the eigendecomposition guarantees common components. 
If our test fails to reject the null hypothesis for the matrices $\Phi^c$, it provides us confidence to conduct the transformation $\bm{z}_{t}^c = V \bm{y}_{t}^c$, $\bm{\mu}_*^c = V \bm{\mu}^c$ with $V \Phi^c V^{-1} = D^c$ diagonal. The lagged cross-dependence cancels out for each coordinate of $\bm{z}_{t}^c$, i.e.
$\mathbb{E}[\bm{z}_t^c] = \bm{\mu}_*^c + D^c \mathbb{E}[\bm{z}_{t-1}^c]$.
Hence, with $\bm{z}^{c}_t$ as the new set of variables, comparison between countries can then be done on each variable separately.


The quarterly data of seasonally adjusted time series span from the first quarter of 1992 to the first quarter of 2020, with length $n+1 = 113$, and are pre-processed by taking the log-difference and standardization. The least square estimators of the coefficient matrices are then obtained, and they follow asymptotic normal distributions with rate $n^{-1/2}$; see Chapter 3 in \cite{Lutkepol2005} for more details on estimating VAR models. We conduct our multi-sample tests \autoref{cor:multi} and \autoref{thm:multi_eig.Est} on those coefficient matrices.

Implementing \autoref{cor:multi}, we get a p-value 0.991 which indicates that we fail to reject $H_0$ concerning all eight countries. Meanwhile \autoref{thm:multi_eig.Est} yields a p-value of 0.045 at the margin of rejection level. 
These results are not surprising based on our simulation study in \autoref{se:sim-multi}. \autoref{fig:multi} indicates that under the null space, the chi-squared test (\autoref{thm:multi_eig.Est}) tends to reject the hypothesis more often than suggested by the nominal test size and the gamma test (\autoref{cor:multi}) tends to reject the hypothesis less often.

\begin{figure}[htb]
    \centering
    \centerline{\input{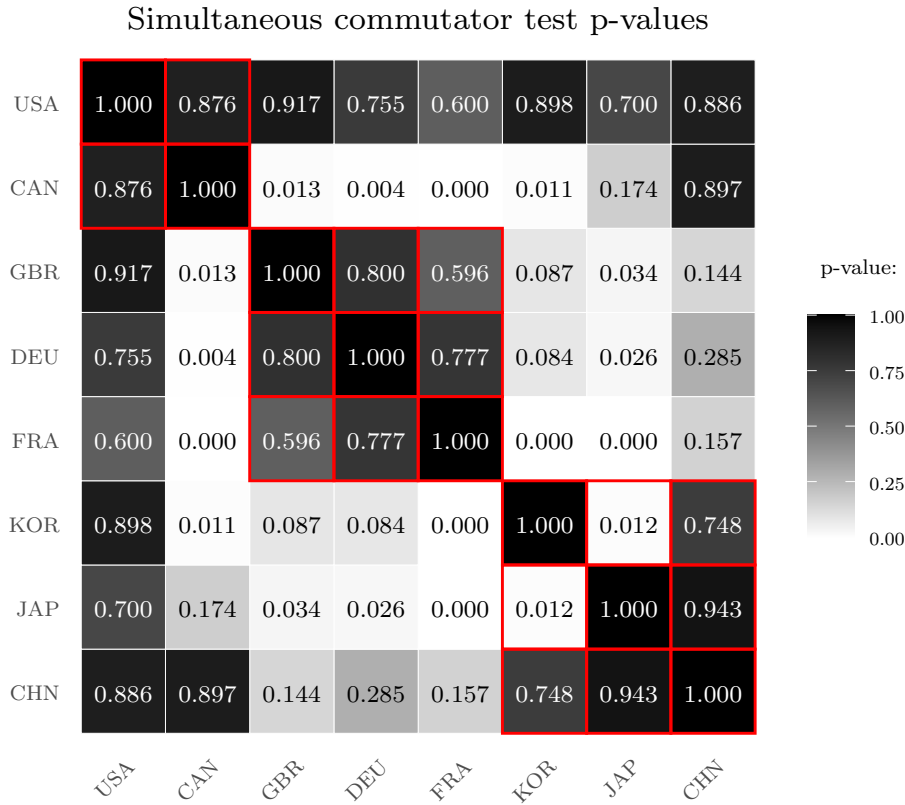}}
    \caption{The heatmap of the p-values of the pairwise commutator-based test.}
    \label{fig:pairP}
\end{figure}

In addition, if we look at the p-value table (\autoref{fig:pairP}) for simultaneous commutator-based tests according to \autoref{thm:comm.Est}, we see that the United States and China both share quite evident similarities with all other countries, and there are reasonable similarity structures within each continental group, except between Korea and Japan. Splitting the eight countries into three continental groups and repeating our test based on \autoref{cor:multi} group-wisely, we successfully get the conclusion that $H_0$ holds within the continental groups. The corresponding p-values are, 0.530 for North America (United States and Canada), 0.978 for Europe (France, Germany, and United Kingdom), and 0.900 for Asia (China, Japan, and Korea). In addition, even the unstable \autoref{thm:multi_eig.Est} gives p-values 0.076 and 0.657 for North America and Europe, respectively.

Since the test statistic in \autoref{cor:multi} gives a relatively large p-value and taking into consideration the test results of our pairwise test as well as the test performances in the simulation study, we may conclude that one fails to reject $H_0$. Our test results provide evidence that it is reasonable to assume that the coefficient matrices share the same eigenvectors. The approximated common left eigenvectors $V$ and the new shared variables are
\begin{equation}
    \bm{z}_t = V  \begin{pmatrix} \mbox{GDP}_t \\ \mbox{M2}_t \\ \mbox{REER}_t \end{pmatrix} = \begin{pmatrix} 0.66 & 0.41 & -0.10 \\
    0.62 & 0.04 & -0.81\\
    0.72 & -1.03 & -0.10\end{pmatrix} \begin{pmatrix} \mbox{GDP}_t \\ \mbox{M2}_t \\ \mbox{REER}_t \end{pmatrix}.\label{eqn:Vinv}
\end{equation}

In addition, we conduct the partial test for one common eigenvector ($k=1$) on all eight coefficient matrices, and get that the p-values equal 0.737 from the chi-squared test \eqref{eqn:partchi} and 0.964 from the gamma approximation \eqref{eqn:part_approx}. In addition, when considering $k = 2$, \eqref{eqn:part_approx} gives the p-value 0.948. Recall that the transformed variables have the following notation $\bm{z}_t^c = (\bm{z}_t[1], \bm{z}_t[2], \bm{z}_t^c[3])$, where the first two variables (without superscript $c$) are shared across all eight countries. The two common variables $\bm{z}_t[1]$ and $\bm{z}_t[2]$ that only depend on itself in expectation are
\begin{align*}
    \bm{z}_t[1] & = 0.67 * \mbox{GDP}_t + 0.34 * \mbox{M2}_t - 0.68 * \mbox{REER}_t,\\
    \bm{z}_t[2] & = 0.58 * \mbox{GDP}_t - 0.93 * \mbox{M2}_t - 0.03 * \mbox{REER}_t.
\end{align*}
Note that $\bm{z}_t[1]$ and $\bm{z}_t[2]$ correspond to the first and third row of $V$ in \eqref{eqn:Vinv} respectively, with a bit rescaling and fluctuations.

\subsection{Stationary distribution of Markov chains}\label{se:Markov}
Consider $p$ recurrent Markov chains $\{X_i = \{X_{i,t}\}_{t = 1}^{n+1} \}_{i = 1}^p$ with time length $n+1$ and labels $X_{i,t}$ from a finite discrete state space $\{1, \dots, d\}$. Within each chain the transition probability matrix $P_i$ can be estimated consistently to follow asymptotic normality. 
For example, $P_i$ can be estimated by
\begin{equation}
    \widehat{P}_i = \big(\widehat{P}_{i,rs}\big)_{d \times d}, ~ \widehat{P}_{i,rs} = \frac{\sum_{t=1}^{n} \mathbb{I}(X_{i,t} = r) \mathbb{I}(X_{i,t+1} = s)}{\sum_{t = 1}^{n} \mathbb{I}(X_{i,t} = r)};\label{eqn:P_est}
\end{equation}
see (3) in \cite{barsotti}.
The estimates $\widehat{P}_i$ in \eqref{eqn:P_est} can be deduced to follow asymptotic normality in the sense of \autoref{ass:normal} with limiting covariance $\Sigma_i$ given by
\begin{equation}
\lim_{n \to \infty} n \cov(\widehat{P}_{i,rs}, \widehat{P}_{i,uv}) = 
\begin{cases}
P_{i,rs} (1 - P_{i,rs})/\bm{\pi}_{i,r}, & r = u, s = v,
\\
-P_{i,rs} P_{i,uv}/\bm{\pi}_{i,r}, & r = u, s \neq v,
\\
0, & \mbox{otherwise,}
\end{cases} \label{eqn:covMarkov}
\end{equation}
where $\bm{\pi}_i = (\bm{\pi}_{i,1}, \dots, \bm{\pi}_{i,d}) \in \mathbb{R}^d$ is the stationary distribution of chain $X_i$; see Lemma 3.1 in \cite{barsotti}. Note that $\bm{\pi}_{i,r}$ is strictly positive for any $i$ and $r$ since all chains are recurrent. In practice, $\bm{\pi}_i$ can be estimated either from $\widehat{P}_i$ or directly from chain $X_i$.

Applying the Perron-Frobenius theorem \citep[Theorem 8.4.4]{Horn}, it is possible to conduct our partial test to check identical stationary distributions if we could find a common non-negative eigenvector. Using the fact that each $P_i$ (and $\widehat{P}_i$) has stationary distribution as an eigenvector corresponding to eigenvalue $1$, we can optimize the common non-negative eigenvector associated with fixed eigenvalue $1$, which must be proportional to the common distribution vector if it exists. We aim to find the non-negative vector $\bm{\widehat{\pi}}$ which optimizes the problem
\begin{multline}\label{eqn:minimize}
    \mbox{minimize } f(\bm{x}) = \sum_{i = 1}^p \bm{x}'(\widehat{P}_i - I_d)(\widehat{P}_i' - I_d)\bm{x},\\
    ~ \mbox{subject to } \sum_{i=1}^d \bm{x}_i = 1, \mbox{ and }\bm{x}_i \ge 0, ~ \forall i = 1, \dots, d.
\end{multline}
It can be solved via quadratic programming with explicitly given constraints.

 \begin{sidewaysfigure}[p] 
     \centering
     \centerline{\input{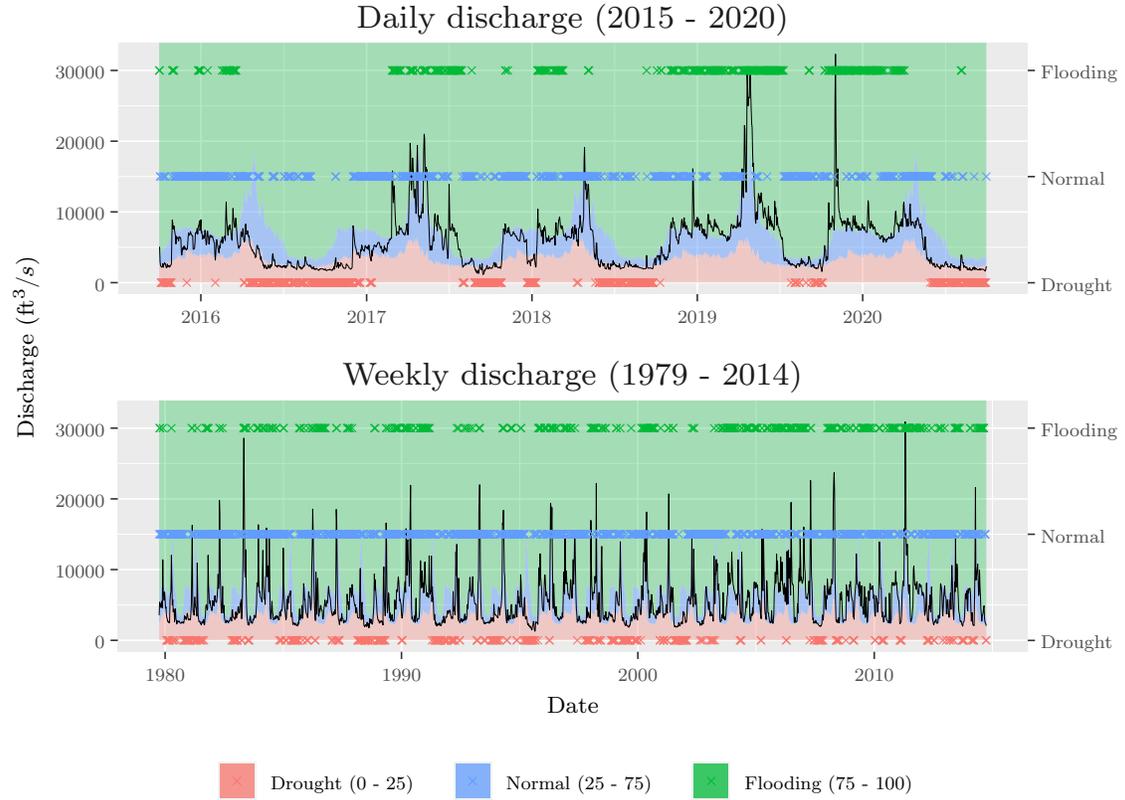}}
     \caption{The discharge dataset for Hudson river with different time resolutions and disjoint time coverage. The unit is $\mbox{ft}^3/s$. The discharges are classified as 'drought' if in the lower region (below 25 percentile), as 'flooding' if in the upper region (above 75 percentile), and as 'normal' if in the middle. In the plot, the (colored) crosses represent the classified states, with axis on the right.}
    \label{fig:stream}
 \end{sidewaysfigure}

\begin{figure}[htbp] 
    \centering
    \centerline{\input{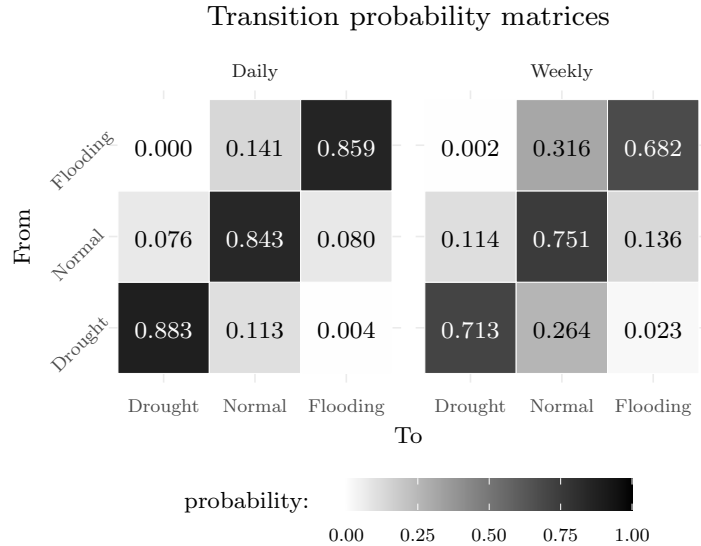}}
    \caption{The 2 transition matrices are estimated according to \eqref{eqn:P_est} and they are statistically different according to asymptotic Wald test. Numerically speaking, at least the diagonal entries of the 2 matrices are obviously unequal.}
   \label{fig:streamTran}
\end{figure}

We consider the streamflow discharge data of Hudson river collected at Fort Edward, NY; see \autoref{fig:stream}. The dataset is available at \cite{usgs}. We use the historical weekly data (from Oct. 1st 1979 to Sept. 30th 2014) and the more recent daily data (from Oct. 1st 2015 to Sept. 30th 2020). Both series have the same length $n = 1827$. We classify the discharge records to 3 levels according to the percentile statistics based on data from Oct. 1st 1977 to Sept. 30th 2019. An observation is referred to as 'drought' if it is below the 25th percentile, as 'flooding' if above the 75th percentile, and as 'normal' otherwise. We assume that this 3-state sequence satisfies the Markov property, and estimate the transition probability matrices (see \autoref{fig:streamTran}) for our partial test. Note that the two time series are disjoint, hence it is reasonable to view these estimators as independent.

We implement the partial test according to both \autoref{thm:part.Est} and \autoref{cor:partial}, with parameter $k = 1$ and $V = \bm{\widehat{\pi}}_i$ given by optimization \eqref{eqn:minimize}. The two tests give the p-values of 0.106 and 0.102 respectively, which indicate that one fails to reject $H_0^{*}$ in \eqref{hyp:part} and may conclude the two series share the same stationary distribution. The estimated common stationary distribution is
$
\mathbb{P}(\text{Drought}) = 0.230, \hspace{0.15cm} \mathbb{P}(\text{Normal}) = 0.519, \hspace{0.15cm} \mathbb{P}(\text{Flooding}) = 0.251.
$
The results indicate that for time series with different time resolutions, even though the transition probability matrices can be different, the stationary distribution might still be verified to be statistically equivalent.

\section{Conclusions}\label{sec:con}
In this work, we focused on general asymmetric matrix-valued population quantities and developed effective tests on simultaneous diagonalizability under both two-sample and multi-sample settings. We also generalized our designs to test on partially shared eigenvectors, with introduction of a supplemental optimization algorithm that retrieves those common eigenvectors. 
Finally, we applied our test to real data examples and successfully revealed interesting structural properties of the underlying models.

In this work, we considered the classical ``fixed $d$, large $n$” regime. However, many contemporary data go beyond the low dimensional setting and require the dimension $d$ to be of the same order as, or possibly even larger than, the sample size $n$.
While the high-dimensional setting goes beyond the scope of this work, we added a discussion and simulation study in \autoref{se:compl3} of the supplementary material to emphasize that the methodology in this paper is not sufficient to do testing on high-dimensional data.

Other possible future directions include adjacency matrices of weighted directed networks which are possibly asymmetric square matrices. Our partial test could be of special interest to test on their leading eigenvectors. 
Combined with the high-dimensional setting, one could even study expanding networks.

In our multi-sample (\autoref{sec:multi}) and partial tests (\autoref{sec:part}), the estimated $\widehat{V}$ is treated as given and deterministic in the test statistics. It might be possible to pursue more rigorous results by deriving the stochastic properties of the optimizer $\widehat{V}$. 

\begin{appendix}
\section{Complementary simulation results} \label{se:compl}
We provide here some empirical results complementary to the numerical analysis presented in the main paper. \autoref{se:compl1} gives empirical sizes and powers for the proposed test, 
\autoref{se:compl2} studies sequential application of our partial tests and \autoref{se:compl3} discusses application to possibly high-dimensional data.

\subsection{Empirical Type I and II errors} \label{se:compl1}
In addition to the p-values in the main paper, we provide here tables with Type I and II errors for our tests to assess their performances. \Cref{ta:errorsTwosample,,ta:errorsMultisample,,ta:errorsPartial} show respectively the errors for the two-sample, multi-sample and partial tests.

\begin{table}[htbp] 
\centering
\begin{tabular}{|c|c|c|c|cc|}
\hline
\multirow{2}{*}{Test Type} & \multirow{2}{*}{\begin{tabular}[c]{@{}c@{}}Statistics\\ Type\end{tabular}} & \multirow{2}{*}{\begin{tabular}[c]{@{}c@{}}Sample\\ Size\end{tabular}} & \multirow{2}{*}{\begin{tabular}[c]{@{}c@{}}Type I\\ Error\end{tabular}} & \multicolumn{2}{c|}{Type II Error} \\ \cline{5-6} 
 &  &  &  & \multicolumn{1}{c|}{SNR=1000} & SNR=10 \\ \hline
\multirow{3}{*}{\begin{tabular}[c]{@{}c@{}}Commutator-based\\ test (\autoref{sec:commute})\end{tabular}} & \multirow{3}{*}{\begin{tabular}[c]{@{}c@{}}Chi-test\\ (\autoref{thm:comm.Est})\end{tabular}} & 50 & 0.218 & \multicolumn{1}{c|}{0.216} & 0.000 \\ \cline{3-6} 
 &  & 250 & 0.056 & \multicolumn{1}{c|}{0.000} & 0.000 \\ \cline{3-6} 
 &  & 1000 & 0.054 & \multicolumn{1}{c|}{0.000} & 0.000 \\ \hline
\multirow{12}{*}{\begin{tabular}[c]{@{}c@{}}LLR test\\ (\autoref{sec:llrt})\end{tabular}} & \multirow{3}{*}{\begin{tabular}[c]{@{}c@{}}Oracle Chi-Test\\ with \eqref{eqn:polyP}\end{tabular}} & 50 & 0.182 & \multicolumn{1}{c|}{0.000} & 0.000 \\ \cline{3-6} 
 &  & 250 & 0.140 & \multicolumn{1}{c|}{0.000} & 0.000 \\ \cline{3-6} 
 &  & 1000 & 0.060 & \multicolumn{1}{c|}{0.000} & 0.000 \\ \cline{2-6} 
 & \multirow{3}{*}{\begin{tabular}[c]{@{}c@{}}Plugin Chi-test\\ with \eqref{eqn:polyP}\end{tabular}} & 50 & 1.000 & \multicolumn{1}{c|}{0.000} & 0.000 \\ \cline{3-6} 
 &  & 250 & 1.000 & \multicolumn{1}{c|}{0.000} & 0.000 \\ \cline{3-6} 
 &  & 1000 & 1.000 & \multicolumn{1}{c|}{0.000} & 0.000 \\ \cline{2-6} 
 & \multirow{3}{*}{\begin{tabular}[c]{@{}c@{}}Oracle Chi-Test\\ with \eqref{eqn:eigvP}\end{tabular}} & 50 & 0.182 & \multicolumn{1}{c|}{0.058} & 0.000 \\ \cline{3-6} 
 &  & 250 & 0.140 & \multicolumn{1}{c|}{0.000} & 0.000 \\ \cline{3-6} 
 &  & 1000 & 0.060 & \multicolumn{1}{c|}{0.000} & 0.000 \\ \cline{2-6} 
 & \multirow{3}{*}{\begin{tabular}[c]{@{}c@{}}Plugin Chi-test\\ with \eqref{eqn:eigvP}\end{tabular}} & 50 & 0.760 & \multicolumn{1}{c|}{0.000} & 0.000 \\ \cline{3-6} 
 &  & 250 & 0.842 & \multicolumn{1}{c|}{0.000} & 0.000 \\ \cline{3-6} 
 &  & 1000 & 0.774 & \multicolumn{1}{c|}{0.000} & 0.000 \\ \hline
\end{tabular}
\caption{Two-sample test results on simulated $\mathcal{M}_2(\rho, 5; 5)$ for $\rho^2 = \frac{1}{\SNR} \in \{ 0, \frac{1}{1000}, \frac{1}{10} \}$.} \label{ta:errorsTwosample}
\end{table}

\begin{table}[htbp]
\centering
\begin{tabular}{|c|c|c|ccc|}
\hline
\multirow{2}{*}{\begin{tabular}[c]{@{}c@{}}Statistics\\ Type\end{tabular}} & \multirow{2}{*}{\begin{tabular}[c]{@{}c@{}}Sample\\ Size\end{tabular}} & \multirow{2}{*}{\begin{tabular}[c]{@{}c@{}}Type I\\ Error\end{tabular}} & \multicolumn{3}{c|}{Type II Error} \\ \cline{4-6} 
 &  &  & \multicolumn{1}{c|}{$\SNR=1000$} & \multicolumn{1}{c|}{$\SNR=100$} & $\SNR=10$ \\ \hline
\multirow{4}{*}{\begin{tabular}[c]{@{}c@{}}Oracle\\ Chi-test\\(\autoref{thm:multi_eig})\end{tabular}} & 100 & 0.230 & \multicolumn{1}{c|}{NA} & \multicolumn{1}{c|}{NA} & NA \\ \cline{2-6} 
 & 1000 & 0.060 & \multicolumn{1}{c|}{NA} & \multicolumn{1}{c|}{NA} & NA \\ \cline{2-6} 
 & 10000 & 0.045 & \multicolumn{1}{c|}{NA} & \multicolumn{1}{c|}{NA} & NA \\ \cline{2-6} 
 & 100000 & 0.070 & \multicolumn{1}{c|}{NA} & \multicolumn{1}{c|}{NA} & NA \\ \hline
\multirow{4}{*}{\begin{tabular}[c]{@{}c@{}}Plugin\\ Chi-test\\(\autoref{thm:multi_eig.Est})\end{tabular}} & 100 & 0.175 & \multicolumn{1}{c|}{0.000} & \multicolumn{1}{c|}{0.000} & 0.000 \\ \cline{2-6} 
 & 1000 & 0.095 & \multicolumn{1}{c|}{0.000} & \multicolumn{1}{c|}{0.000} & 0.000 \\ \cline{2-6} 
 & 10000 & 0.090 & \multicolumn{1}{c|}{0.000} & \multicolumn{1}{c|}{0.000} & 0.000 \\ \cline{2-6} 
 & 100000 & 0.075 & \multicolumn{1}{c|}{0.000} & \multicolumn{1}{c|}{0.000} & 0.000 \\ \hline
\multirow{4}{*}{\begin{tabular}[c]{@{}c@{}}Plugin\\ Gamma-test\\(\autoref{cor:multi})\end{tabular}} & 100 & 0.015 & \multicolumn{1}{c|}{0.005} & \multicolumn{1}{c|}{0.000} & 0.000 \\ \cline{2-6} 
 & 1000 & 0.025 & \multicolumn{1}{c|}{0.000} & \multicolumn{1}{c|}{0.000} & 0.000 \\ \cline{2-6} 
 & 10000 & 0.005 & \multicolumn{1}{c|}{0.000} & \multicolumn{1}{c|}{0.000} & 0.000 \\ \cline{2-6} 
 & 100000 & 0.015 & \multicolumn{1}{c|}{0.000} & \multicolumn{1}{c|}{0.000} & 0.000 \\ \hline
\end{tabular}
\caption{Multi-sample test results on simulated $\mathcal{M}_8(\rho, 4; 4)$ for $\rho^2 = \frac{1}{\SNR} \in \{ 0, \frac{1}{1000}, \frac{1}{100}\, \frac{1}{10} \}$.} \label{ta:errorsMultisample}
\end{table}

\begin{table}[htbp]
\centering
\begin{tabular}{|c|c|c|ccc|}
\hline
\multirow{2}{*}{Statistics Type} & \multirow{2}{*}{Sample Size} & \multirow{2}{*}{Type I Error} & \multicolumn{3}{c|}{Type II Error} \\ \cline{4-6} 
 &  &  & \multicolumn{1}{c|}{$\SNR=1000$} & \multicolumn{1}{c|}{$\SNR=100$} & $\SNR=10$ \\ \hline
\multirow{3}{*}{\begin{tabular}[c]{@{}c@{}}Chi-test\\(\autoref{thm:part.Est})\end{tabular}} & 100 & 0.020 & \multicolumn{1}{c|}{0.000} & \multicolumn{1}{c|}{0.000} & 0.000 \\ \cline{2-6} 
 & 1000 & 0.020 & \multicolumn{1}{c|}{0.000} & \multicolumn{1}{c|}{0.000} & 0.000 \\ \cline{2-6} 
 & 10000 & 0.025 & \multicolumn{1}{c|}{0.000} & \multicolumn{1}{c|}{0.000} & 0.000 \\ \hline
\multirow{3}{*}{\begin{tabular}[c]{@{}c@{}}Gamma-test\\(\autoref{cor:partial})\end{tabular}} & 100 & 0.010 & \multicolumn{1}{c|}{0.000} & \multicolumn{1}{c|}{0.000} & 0.000 \\ \cline{2-6} 
 & 1000 & 0.015 & \multicolumn{1}{c|}{0.000} & \multicolumn{1}{c|}{0.000} & 0.000 \\ \cline{2-6} 
 & 10000 & 0.015 & \multicolumn{1}{c|}{0.000} & \multicolumn{1}{c|}{0.000} & 0.000 \\ \hline
\end{tabular}
\caption{Partial test results on simulated $\mathcal{M}_8(\rho, 2; 4)$ for $\rho = \frac{1}{\SNR} \in \{ 0, \frac{1}{1000}, \frac{1}{100}\, \frac{1}{10} \}$.} \label{ta:errorsPartial}
\end{table}

\subsection{Sequential application of partial tests} \label{se:compl2}
As pointed out in \autoref{sec:part}, we assume that the number of partial common eigenvectors in known. Since this assumption is not feasible in practice, we propose a sequential testing procedure. The hypothesis testing problem \eqref{hyp:part} can be stated for $k \in \{1,\dots,d\}$. The sequential testing starts with $k = d$, then $k = d-1$ and so on, till the null hypothesis is not rejected.
The performance of this procedure is accessed through a simulation study in \autoref{ta:sequential}.

\begin{table}[htbp]
\centering
\begin{tabular}{|c|c|ccc|}
\hline
\multirow{2}{*}{Statistics Type} & \multirow{2}{*}{Sample Size} & \multicolumn{3}{c|}{Rejection Rate} \\ \cline{3-5} 
 &  & \multicolumn{1}{c|}{$k=2$} & \multicolumn{1}{c|}{$k=3$} & $k=4$ \\ \hline
\multirow{3}{*}{\begin{tabular}[c]{@{}c@{}}Chi-test\\(\autoref{thm:part.Est})\end{tabular}} & 100 & \multicolumn{1}{c|}{0.020} & \multicolumn{1}{c|}{1.000} & 1.000 \\ \cline{2-5} 
 & 1000 & \multicolumn{1}{c|}{0.020} & \multicolumn{1}{c|}{1.000} & 1.000 \\ \cline{2-5} 
 & 10000 & \multicolumn{1}{c|}{0.025} & \multicolumn{1}{c|}{1.000} & 1.000 \\ \hline
\multirow{3}{*}{\begin{tabular}[c]{@{}c@{}}Gamma-test\\(\autoref{cor:partial})\end{tabular}} & 100 & \multicolumn{1}{c|}{0.010} & \multicolumn{1}{c|}{1.000} & 1.000 \\ \cline{2-5} 
 & 1000 & \multicolumn{1}{c|}{0.015} & \multicolumn{1}{c|}{1.000} & 1.000 \\ \cline{2-5} 
 & 10000 & \multicolumn{1}{c|}{0.015} & \multicolumn{1}{c|}{1.000} & 1.000 \\ \hline
\end{tabular}
\caption{Partial test results on simulated $\mathcal{M}_8(0, 2; 4)$ and potentially mis-specified $k \in \{2,3,4\}$.} \label{ta:sequential}
\end{table}

\subsection{High-dimensional data} \label{se:compl3}

In this work, we consider the classical ``fixed $d$, large $n$” regime. However, many contemporary data go beyond the low dimensional setting and require the dimension $d$ to be of the same order as, or possibly even larger than, the sample size $n$.
While the high-dimensional setting goes beyond the scope of this work, we would like to point out why our methodology is not sufficient to do testing on high-dimensional data.

\autoref{ta:differentdimensions} presents the empirical rejection rates and average degrees of freedom for the two-sample commutator-based test in \Cref{thm:comm,,thm:comm.Est}, considering different sample sizes $n=50, 100, 500$ and letting $d$ grow. We present results assuming that the limiting covariance matrix in \eqref{eq:comm.asym} is estimated and known. The existence of a consistent estimator is stated in \autoref{ass:covConsistent} and makes our procedure feasible in practice.

The results in \autoref{ta:differentdimensions} show that the classical theory suffers a $\alpha$ test size much higher than the nominal test level once we consider high-dimensional data and estimate the limiting covariance matrix. Intuitively, the results are expected to break down once the sample size does not satisfy $n > r_{1}(d^2 + d^2)$. This can be easily seen by counting the degrees of freedom required to specify a rank-$r_{1}$ matrix of size $d^2 \times d^2$. Roughly speaking, we need $r_{1}$ numbers to specify the matrix's singular values, and $r_{1}d^2$ and $r_{1}d^2$ numbers to specify its left and right singular vectors.

The $\alpha$ test size much higher than the nominal test level is also due to \autoref{ass:covConsistent} no longer being satisfied in a high-dimensional regime. In particular, the difference between estimator and true matrix is incorrectly normalized once the dimension grows with the sample size. It is expected to require results from random matrix theory to get convergence under suitable assumptions on the ratio between $d$ and $n$.

\begin{sidewaystable}[htbp]
\centering
  \small
\begin{tabular}{|c|cccc|cccc|cccc|}
\hline
\multirow{3}{*}{$d$} & \multicolumn{4}{c|}{Sample Size=50} & \multicolumn{4}{c|}{Sample Size=100} & \multicolumn{4}{c|}{Sample Size = 500} \\ \cline{2-13} 
 & \multicolumn{2}{c|}{Empirical Cov} & \multicolumn{2}{c|}{True Cov} & \multicolumn{2}{c|}{Empirical Cov} & \multicolumn{2}{c|}{True Cov} & \multicolumn{2}{c|}{Empirical Cov} & \multicolumn{2}{c|}{True Cov} \\ \cline{2-13} 
 & \multicolumn{1}{c|}{Size} & \multicolumn{1}{c|}{Avg DF} & \multicolumn{1}{c|}{Size} & Avg DF & \multicolumn{1}{c|}{Size} & \multicolumn{1}{c|}{Avg DF} & \multicolumn{1}{c|}{Size} & Avg DF & \multicolumn{1}{c|}{Size} & \multicolumn{1}{c|}{Avg DF} & \multicolumn{1}{c|}{Size} & Avg DF \\ \hline
2 & \multicolumn{1}{c|}{0.036} & \multicolumn{1}{c|}{2.05} & \multicolumn{1}{c|}{0.022} & 2.04 & \multicolumn{1}{c|}{0.054} & \multicolumn{1}{c|}{2.00} & \multicolumn{1}{c|}{0.058} & 2.00 & \multicolumn{1}{c|}{0.044} & \multicolumn{1}{c|}{2.00} & \multicolumn{1}{c|}{0.044} & 2.00 \\ \hline
3 & \multicolumn{1}{c|}{0.066} & \multicolumn{1}{c|}{6.10} & \multicolumn{1}{c|}{0.034} & 6.13 & \multicolumn{1}{c|}{0.074} & \multicolumn{1}{c|}{6.02} & \multicolumn{1}{c|}{0.054} & 6.02 & \multicolumn{1}{c|}{0.062} & \multicolumn{1}{c|}{6.00} & \multicolumn{1}{c|}{0.060} & 6.00 \\ \hline
4 & \multicolumn{1}{c|}{0.088} & \multicolumn{1}{c|}{12.33} & \multicolumn{1}{c|}{0.014} & 12.38 & \multicolumn{1}{c|}{0.092} & \multicolumn{1}{c|}{12.06} & \multicolumn{1}{c|}{0.036} & 12.06 & \multicolumn{1}{c|}{0.066} & \multicolumn{1}{c|}{12.00} & \multicolumn{1}{c|}{0.054} & 12.00 \\ \hline
5 & \multicolumn{1}{c|}{0.186} & \multicolumn{1}{c|}{20.15} & \multicolumn{1}{c|}{0.018} & 20.42 & \multicolumn{1}{c|}{0.116} & \multicolumn{1}{c|}{20.05} & \multicolumn{1}{c|}{0.050} & 20.08 & \multicolumn{1}{c|}{0.062} & \multicolumn{1}{c|}{20.00} & \multicolumn{1}{c|}{0.056} & 20.00 \\ \hline
6 & \multicolumn{1}{c|}{0.358} & \multicolumn{1}{c|}{29.75} & \multicolumn{1}{c|}{0.020} & 30.65 & \multicolumn{1}{c|}{0.176} & \multicolumn{1}{c|}{29.97} & \multicolumn{1}{c|}{0.020} & 30.09 & \multicolumn{1}{c|}{0.058} & \multicolumn{1}{c|}{30.00} & \multicolumn{1}{c|}{0.046} & 30.00 \\ \hline
7 & \multicolumn{1}{c|}{0.698} & \multicolumn{1}{c|}{40.85} & \multicolumn{1}{c|}{0.018} & 42.79 & \multicolumn{1}{c|}{0.310} & \multicolumn{1}{c|}{40.46} & \multicolumn{1}{c|}{0.026} & 40.97 & \multicolumn{1}{c|}{0.092} & \multicolumn{1}{c|}{40.00} & \multicolumn{1}{c|}{0.050} & 40.00 \\ \hline
8 & \multicolumn{1}{c|}{0.960} & \multicolumn{1}{c|}{55.82} & \multicolumn{1}{c|}{0.024} & 57.78 & \multicolumn{1}{c|}{0.548} & \multicolumn{1}{c|}{56.04} & \multicolumn{1}{c|}{0.034} & 56.30 & \multicolumn{1}{c|}{0.092} & \multicolumn{1}{c|}{56.00} & \multicolumn{1}{c|}{0.044} & 56.00 \\ \hline
9 & \multicolumn{1}{c|}{1.000} & \multicolumn{1}{c|}{69.11} & \multicolumn{1}{c|}{0.014} & 74.19 & \multicolumn{1}{c|}{0.778} & \multicolumn{1}{c|}{70.54} & \multicolumn{1}{c|}{0.044} & 71.97 & \multicolumn{1}{c|}{0.098} & \multicolumn{1}{c|}{70.44} & \multicolumn{1}{c|}{0.052} & 70.57 \\ \hline
10 & \multicolumn{1}{c|}{1.000} & \multicolumn{1}{c|}{86.27} & \multicolumn{1}{c|}{0.020} & 92.94 & \multicolumn{1}{c|}{0.962} & \multicolumn{1}{c|}{89.62} & \multicolumn{1}{c|}{0.034} & 90.74 & \multicolumn{1}{c|}{0.162} & \multicolumn{1}{c|}{90.00} & \multicolumn{1}{c|}{0.054} & 90.00 \\ \hline
11 & \multicolumn{1}{c|}{1.000} & \multicolumn{1}{c|}{96.93} & \multicolumn{1}{c|}{0.024} & 114.55 & \multicolumn{1}{c|}{0.996} & \multicolumn{1}{c|}{108.25} & \multicolumn{1}{c|}{0.030} & 111.15 & \multicolumn{1}{c|}{0.238} & \multicolumn{1}{c|}{108.79} & \multicolumn{1}{c|}{0.050} & 109.15 \\ \hline
12 & \multicolumn{1}{c|}{1.000} & \multicolumn{1}{c|}{98.00} & \multicolumn{1}{c|}{0.030} & 136.88 & \multicolumn{1}{c|}{1.000} & \multicolumn{1}{c|}{131.96} & \multicolumn{1}{c|}{0.042} & 133.48 & \multicolumn{1}{c|}{0.340} & \multicolumn{1}{c|}{132.00} & \multicolumn{1}{c|}{0.056} & 132.00 \\ \hline
13 & \multicolumn{1}{c|}{1.000} & \multicolumn{1}{c|}{98.00} & \multicolumn{1}{c|}{0.022} & 161.63 & \multicolumn{1}{c|}{1.000} & \multicolumn{1}{c|}{153.87} & \multicolumn{1}{c|}{0.030} & 157.93 & \multicolumn{1}{c|}{0.468} & \multicolumn{1}{c|}{156.00} & \multicolumn{1}{c|}{0.044} & 156.00 \\ \hline
14 & \multicolumn{1}{c|}{1.000} & \multicolumn{1}{c|}{98.00} & \multicolumn{1}{c|}{0.028} & 188.91 & \multicolumn{1}{c|}{1.000} & \multicolumn{1}{c|}{174.85} & \multicolumn{1}{c|}{0.026} & 184.56 & \multicolumn{1}{c|}{0.622} & \multicolumn{1}{c|}{182.00} & \multicolumn{1}{c|}{0.044} & 182.00 \\ \hline
15 & \multicolumn{1}{c|}{1.000} & \multicolumn{1}{c|}{98.00} & \multicolumn{1}{c|}{0.040} & 217.63 & \multicolumn{1}{c|}{1.000} & \multicolumn{1}{c|}{193.89} & \multicolumn{1}{c|}{0.042} & 212.96 & \multicolumn{1}{c|}{0.774} & \multicolumn{1}{c|}{209.88} & \multicolumn{1}{c|}{0.052} & 209.99 \\ \hline
16 & \multicolumn{1}{c|}{0.998} & \multicolumn{1}{c|}{98.00} & \multicolumn{1}{c|}{0.042} & 248.27 & \multicolumn{1}{c|}{1.000} & \multicolumn{1}{c|}{198.00} & \multicolumn{1}{c|}{0.034} & 243.25 & \multicolumn{1}{c|}{0.874} & \multicolumn{1}{c|}{239.89} & \multicolumn{1}{c|}{0.042} & 240.00 \\ \hline
17 & \multicolumn{1}{c|}{0.998} & \multicolumn{1}{c|}{98.00} & \multicolumn{1}{c|}{0.050} & 281.03 & \multicolumn{1}{c|}{1.000} & \multicolumn{1}{c|}{198.00} & \multicolumn{1}{c|}{0.042} & 275.62 & \multicolumn{1}{c|}{0.974} & \multicolumn{1}{c|}{270.00} & \multicolumn{1}{c|}{0.046} & 270.14 \\ \hline
18 & \multicolumn{1}{c|}{0.978} & \multicolumn{1}{c|}{98.00} & \multicolumn{1}{c|}{0.050} & 315.85 & \multicolumn{1}{c|}{1.000} & \multicolumn{1}{c|}{198.00} & \multicolumn{1}{c|}{0.036} & 310.37 & \multicolumn{1}{c|}{0.986} & \multicolumn{1}{c|}{303.18} & \multicolumn{1}{c|}{0.056} & 304.74 \\ \hline
19 & \multicolumn{1}{c|}{0.994} & \multicolumn{1}{c|}{98.00} & \multicolumn{1}{c|}{0.046} & 352.90 & \multicolumn{1}{c|}{1.000} & \multicolumn{1}{c|}{198.00} & \multicolumn{1}{c|}{0.050} & 347.06 & \multicolumn{1}{c|}{0.998} & \multicolumn{1}{c|}{340.60} & \multicolumn{1}{c|}{0.044} & 341.50 \\ \hline
20 & \multicolumn{1}{c|}{0.994} & \multicolumn{1}{c|}{98.00} & \multicolumn{1}{c|}{0.056} & 391.32 & \multicolumn{1}{c|}{1.000} & \multicolumn{1}{c|}{198.00} & \multicolumn{1}{c|}{0.046} & 384.37 & \multicolumn{1}{c|}{1.000} & \multicolumn{1}{c|}{373.18} & \multicolumn{1}{c|}{0.044} & 375.67 \\ \hline
21 & \multicolumn{1}{c|}{0.460} & \multicolumn{1}{c|}{98.00} & \multicolumn{1}{c|}{0.054} & 433.27 & \multicolumn{1}{c|}{1.000} & \multicolumn{1}{c|}{198.00} & \multicolumn{1}{c|}{0.040} & 426.84 & \multicolumn{1}{c|}{1.000} & \multicolumn{1}{c|}{418.44} & \multicolumn{1}{c|}{0.044} & 419.91 \\ \hline
22 & \multicolumn{1}{c|}{0.982} & \multicolumn{1}{c|}{98.00} & \multicolumn{1}{c|}{0.044} & 475.39 & \multicolumn{1}{c|}{1.000} & \multicolumn{1}{c|}{198.00} & \multicolumn{1}{c|}{0.040} & 468.52 & \multicolumn{1}{c|}{1.000} & \multicolumn{1}{c|}{455.78} & \multicolumn{1}{c|}{0.054} & 459.21 \\ \hline
23 & \multicolumn{1}{c|}{0.986} & \multicolumn{1}{c|}{98.00} & \multicolumn{1}{c|}{0.044} & 520.58 & \multicolumn{1}{c|}{1.000} & \multicolumn{1}{c|}{198.00} & \multicolumn{1}{c|}{0.054} & 513.53 & \multicolumn{1}{c|}{1.000} & \multicolumn{1}{c|}{499.67} & \multicolumn{1}{c|}{0.042} & 503.17 \\ \hline
24 & \multicolumn{1}{c|}{0.284} & \multicolumn{1}{c|}{98.00} & \multicolumn{1}{c|}{0.032} & 567.74 & \multicolumn{1}{c|}{1.000} & \multicolumn{1}{c|}{198.00} & \multicolumn{1}{c|}{0.030} & 560.31 & \multicolumn{1}{c|}{1.000} & \multicolumn{1}{c|}{544.63} & \multicolumn{1}{c|}{0.040} & 551.13 \\ \hline
25 & \multicolumn{1}{c|}{0.702} & \multicolumn{1}{c|}{98.00} & \multicolumn{1}{c|}{0.056} & 617.31 & \multicolumn{1}{c|}{1.000} & \multicolumn{1}{c|}{198.00} & \multicolumn{1}{c|}{0.046} & 609.14 & \multicolumn{1}{c|}{1.000} & \multicolumn{1}{c|}{590.03} & \multicolumn{1}{c|}{0.054} & 595.95 \\ \hline
\end{tabular}
\caption{Two-sample commutator-based test results on simulated $\mathcal{M}_2(0, d; d)$ for dimensions $d \in \{ 2, \dots, 25 \}$.} \label{ta:differentdimensions}
\label{ta:highd}
\end{sidewaystable}

\section{Proofs}\label{app:profs}

We provide here the proofs of most theoretical results except \Cref{thm:part.Est,,thm:multi_eig.Est,,thm:LLRest,,thm:comm.Est}. The proofs of those results can be found in \autoref{app:est} since they are based on a generic result.

\begin{proof}[Proof of \autoref{lem:ginverse}]
Under the conditions that given $\varepsilon > 0$ and $\varepsilon$ is not an eigenvalue of $\Sigma$, the mappings $\Sigma \mapsto \Sigma(\varepsilon)$, $\Sigma \mapsto \Sigma^+(\varepsilon)$ and $\Sigma \mapsto \rank(\Sigma; \varepsilon)$ are all at least locally continuous at $\Sigma$. Hence \eqref{eq:wald.cmt} follows by the continuous mapping theorem.
\end{proof}

\begin{proof}[Proof of \autoref{thm:comm}] \label{app:prop1}
Introduce the two vectors
$$
\bm{a}_0 = \big(\Vector(A_1)', \Vector(A_2)'\big)', ~ \bm{m}_0 = \big(\Vector(M_1)', \Vector(M_2)'\big)'.
$$
By \autoref{ass:normal},
\begin{equation} \label{eq:proof3eq1}
c(n) (\bm{a}_0 - \bm{m}_0) \xrightarrow{\mathcal{D}} \mathcal{N}(0, \Sigma_0)
\end{equation}
with $\Sigma_0 = \blkdiag(\Sigma_1, \Sigma_2)$. Recall that $[M_{1},M_{2} ]= M_{1}M_{2}-M_{2}M_{1}$ and define the function $g: \mathbb{R}^{2d^2} \to \mathbb{R}^{d^2}$ such that for $\bm{x} = \big(\bm{x}_1', \bm{x}_2'\big)'$ with $\bm{x}_1, \bm{x}_2 \in \mathbb{R}^{d^2}$,
$$
g(\bm{x}) = \Vector[\mat_d(\bm{x}_1), \mat_d(\bm{x}_2)].
$$
Under $H_0$, we know that $g(\bm{m}_0) = \bm{0}$ and $g(\bm{a}_0) = \bm{\eta}_n = \Vector[A_{1},A_{2} ]$, hence the asymptotic distribution of $\bm{\eta}_n$ can be derived via delta method.

Define
$$
\nabla_g := \nabla g(\bm{m}_0) = \begin{pmatrix}\Lambda(M_2)\\ \Lambda(M_1)\end{pmatrix}.
$$
Then, by delta method and \eqref{eq:proof3eq1},

$$
c(n) \big( g(\bm{a}_0) - g(\bm{m}_0) \big) \xrightarrow{\mathcal{D}} \mathcal{N}(0, \Sigma_\eta),
$$
where $\Sigma_\eta = \nabla_g' \Sigma_0 \nabla_g$. Then \eqref{eqn:gamma2} follows directly by the continuous mapping theorem.
\end{proof}

\begin{proof}[Proof of \autoref{thm:comm.Est}] 
This is a direct corollary from \autoref{thm:generalwald} in \autoref{app:est}.
\end{proof}

Before we prove \autoref{thm:ahat}, we introduce the following lemma.

\begin{lem}\label{lem:poly}
Suppose a matrix $C \in \R^{d \times d}$ has distinct real non-zero eigenvalues. Then any square matrix with the same eigenvectors as $C$ can be expressed by polynomials of $C$ with order less than $d$.
\end{lem}

\begin{proof}[Proof of \autoref{lem:poly}] \label{app:lempoly}
Assume $C = V D_C U$ where $V = (v_1, \dots, v_d)$ is the eigenvector matrix, $U = (u_1, \dots, u_d)' = V^{-1}$, and $D_C$ is the diagonal matrix whose diagonal elements are the corresponding eigenvalues. Then the matrices that also have $V$ as the eigenvector matrix form the following linear space:
$$
\mathfrak{S} = \{X: X = V (\sum_{i=1}^d \bm{a}_i E_i) U = \sum_{i=1}^d \bm{a}_i v_i u_i', ~ \bm{a} = (\bm{a}_1, \dots, \bm{a}_d) \in \mathbb{R}^{d}\},
$$
where the matrix $E_i \in \mathbb{R}^{d \times d}$ has only one non-zero element $1$ at the $i$th diagonal entry. Then the space $\mathfrak{S}$ has dimension $d$ and linear basis $\{v_i u_i'\}_{i=1}^d$.

Since $C$ has distinct real non-zero eigenvalues, according to Cayley-Hamilton theorem \citep[Theorem 2.4.3.2]{Horn}, the characteristic polynomial and the minimal polynomial of $C$ coincide and have degree $d$. Hence the matrices $\{C^0 = I_d, C^1, \dots, C^{d-1}\}$ form an independent basis and the polynomial space
$$
\mathfrak{S}_P = \{X: X = \sum_{i=1}^d \bm{b}_i C^{i-1}, ~ \bm{b} = (\bm{b}_1, \dots, \bm{b}_d) \in \mathbb{R}^{d}\}
$$
has dimension $d$ as well. In addition, since $\mathfrak{S}_P \subset \mathfrak{S}$ and both spaces have dimension $d$, it follows $\mathfrak{S}_P = \mathfrak{S}$.
\end{proof}

\begin{proof}[Proof of \autoref{thm:ahat}]

\autoref{lem:poly} and its proof immediately imply that under the null hypothesis $H_0$, there exist coordinate vectors $\bm{x}_1, \bm{x}_2 \in \R^d$, such that
\begin{equation}
    \Vector(M_1) = P_2 \bm{x}_1, ~ \Vector(M_2) = P_1 \bm{x}_2,\label{eq:x12}
\end{equation}
with matrix $P_i$, $i=1, 2$, defined by either \eqref{eqn:polyP} or \eqref{eqn:eigvP}.

Then the maximization problem in \eqref{eqn:Mest} can be transformed to optimizing with respect to free vectors $\bm{x}_1$ and $\bm{x}_2$ in $\mathbb{R}^d$. By setting the first-order derivative of function $L$ in \eqref{eqn:loglike} to zero, the estimators can then be solved explicitly as
\begin{align*}
    & \Vector(\widehat{A}_1) = P_2 (P_2' \Sigma_1^+ P_2)^+ P_2' \Sigma_1^+ \Vector(A_1),\\
    & \Vector(\widehat{A}_2) = P_1 (P_1' \Sigma_2^+ P_1)^+ P_1' \Sigma_2^+ \Vector(A_2).
\end{align*}
\end{proof}

\begin{proof}[Proof of \autoref{thm:LLR}]
Since $\Sigma_1$ is a positive semidefinite matrix, we can find the low-rank square root $\Sigma_1^{+/2} \in \mathbb{R}^{r \times d^2}$ of its general inverse, such that $\Sigma_1^+ = (\Sigma_1^{+/2})' \Sigma_1^{+/2}$, where $r$ is the rank of $\Sigma_1$. Then, by \autoref{ass:normal},
$$
c(n) \Sigma_1^{+/2} \Vector(A_1 - M_1) \xrightarrow{\mathcal{D}} \mathcal{N}(0, I_{r}).
$$
Set $G = I_{r} - \Sigma_1^{+/2} P_2 (P_2' \Sigma_1^+ P_2)^+ P_2' (\Sigma_1^{+/2})'$, then under $H_0$, the first summand of $\Gamma_2$ in \eqref{eqn:gamma1} is
\begin{align}
    &c^2(n) \Vector (A_1)' Q_{1,2} \Vector(A_1) \nonumber\\
    & = c^2(n) \Vector (A_1)' \Big(\Sigma_1^+ - \Sigma_1^+ P_2 (P_2' \Sigma_1^+ P_2)^+ P_2' \Sigma_1^+\Big) \Vector(A_1) \nonumber\\
    & = c^2(n) \big(\Sigma_1^{+/2} \Vector(A_1)\big)' G \big(\Sigma_1^{+/2} \Vector(A_1)\big) \notag \\
    & = c^2(n) \big(\Sigma_1^{+/2} \Vector(A_1 - M_1)\big)' G \big(\Sigma_1^{+/2} \Vector(A_1 - M_1)\big).\label{eqn:addzeros}
\end{align}
The equality in \eqref{eqn:addzeros} uses the fact that $G \Sigma_1^{+/2} \Vector(M_1) = G \Sigma_1^{+/2} P_2 \bm{x}_1 = 0$. In addition, the matrices $G$ and $I_{r} - G$ are both projection matrices, and $I_{r} - G$ projects matrices onto the column space of $\Sigma_1^{+/2} P_2$, hence $\rank(I_{r} - G) = \rank(P_2' \Sigma_1^+ P_2)$ and $\rank(G) = r - \rank(P_2' \Sigma_1^+ P_2) = r_{1,2}$ with $r_{1,2}$ defined in \autoref{thm:LLR}. Then the first summand in \eqref{eqn:gamma1} satisfies
$$
c^2(n) \big(\Sigma_1^{+/2} \Vector(A_1 - M_1)\big)' G \big(\Sigma_1^{+/2} \Vector(A_1 - M_1)\big) \xrightarrow{\mathcal{D}} \chi^2(r_{1,2}).
$$
Similarly, the second summand in $\Gamma_2$, $c^2(n) \Vector (A_2)' Q_{2,1} \Vector(A_2)$, converges to a chi-square distribution with $r_{2,1}$ degrees of freedom, and since the two summands are independent, the result \eqref{eqn:projTest} follows.
\end{proof}

\begin{proof}[Proof of \autoref{prop:alt.behave}]
Denote
\begin{equation}\label{eqn:sigmaHalf}
    \widehat{\Sigma}_1^+(\varepsilon) = (\widehat{\Sigma}_{1,\varepsilon}^{+/2})' \widehat{\Sigma}_{1, \varepsilon}^{+/2}, \hspace{0.3cm} \text{ with } \hspace{0.3cm} \widehat{\Sigma}_{1,\varepsilon}^{+/2} \in \mathbb{R}^{\widehat{r} \times d^2}, ~ \widehat{r}(\varepsilon) = \rank(\widehat{\Sigma}_1; \varepsilon)
\end{equation}
and
$$
G[\varepsilon] = I_{\widehat{r}(\varepsilon)} - \widehat{\Sigma}_{1,\varepsilon}^{+/2} P_2 \big(P_2' \widehat{\Sigma}_1^+(\varepsilon) P_2\big)^+ P_2' (\widehat{\Sigma}_{1,\varepsilon}^{+/2})'.
$$
Then, the first summand of $\Gamma_2^\#(\varepsilon)$ can be written as
\begin{equation} \label{eqn:alt.zeros}
\begin{aligned}
    &c^2(n) \big(\widehat{\Sigma}_{1,\varepsilon}^{+/2} \Vector(A_1)\big)' G[\varepsilon] \big(\widehat{\Sigma}_{1,\varepsilon}^{+/2} \Vector(A_1)\big)\\
    &=c^2(n) \big(\widehat{\Sigma}_{1,\varepsilon}^{+/2} \Vector(A_1 - M_1)\big)' G[\varepsilon] \big(\widehat{\Sigma}_{1,\varepsilon}^{+/2} \Vector(A_1 - M_1)\big)\\
    & \hspace{1cm}
    + 2 c^2(n) \big(\widehat{\Sigma}_{1,\varepsilon}^{+/2} \Vector(A_1 - M_1)\big)' G[\varepsilon] \widehat{\Sigma}_{1,\varepsilon}^{+/2} \bm{m}_1\\
    & \hspace{2cm}+ c^2(n) \bm{m}_1' (\widehat{\Sigma}_{1,\varepsilon}^{+/2})' G[\varepsilon]
     \widehat{\Sigma}_{1,\varepsilon}^{+/2} \bm{m}_1.
\end{aligned}
\end{equation}
The equality in \eqref{eqn:alt.zeros} uses the fact that
$$
G[\varepsilon] \widehat{\Sigma}_{1,\varepsilon}^{+/2} \Vector(M_1) = G[\varepsilon] \widehat{\Sigma}_{1,\varepsilon}^{+/2} \bm{m}_1.
$$
In addition, it can be verified that $\bm{m}_1' \widehat{\Sigma}_1^+(\varepsilon) P_2 = 0$, then
$$
\bm{m}_1' (\widehat{\Sigma}_{1,\varepsilon}^{+/2})' G[\varepsilon] \widehat{\Sigma}_{1,\varepsilon}^{+/2} \bm{m}_1 = \bm{m}_1' \widehat{\Sigma}_1^+(\varepsilon) \bm{m}_1 > 0,
$$
since $\widehat{\Sigma}_1^+(\varepsilon) \bm{m}_1 \neq 0$. Hence, the deterministic leading term in \eqref{eqn:alt.zeros}, $c^2(n) \bm{m}_1' \widehat{\Sigma}_1^+(\varepsilon) \bm{m}_1$, goes to infinity as $c(n)$ goes to infinity. Similar arguments can be used for the second summand of $\Gamma_2^\#(\varepsilon)$ and the proposition is proved.
\end{proof}


\begin{proof}[Proof of \autoref{lem:asyerror}] \label{app:error}

For $\Delta_\varepsilon$ in \eqref{eqn:deltaeps}, use the notation of $\widehat{\Sigma}_{1,\varepsilon}^{+/2}$ introduced in \eqref{eqn:sigmaHalf} to simplify 
\begin{equation} 
\widehat{Q}_{1,2}[\varepsilon] - \widehat{\mathcal{Q}}_{1,2}[\varepsilon]
= \big(\widehat{\Sigma}_{1,\varepsilon}^{+/2}\big)' (\Delta \widehat{Q}_\varepsilon) \widehat{\Sigma}_{1,\varepsilon}^{+/2},
\end{equation}
where
$$
\Delta \widehat{Q}_\varepsilon = \widehat{Q}_\varepsilon(P_2) - \widehat{Q}_\varepsilon(\widehat{P}_2),
$$
with
$\widehat{Q}_\varepsilon: \mathbb{R}^{d^2 \times d} \to \mathbb{R}^{\widehat{r}(\varepsilon) \times \widehat{r}(\varepsilon)}$ a matrix-valued projection function given by
$$
\widehat{Q}_\varepsilon(P) := \widehat{\Sigma}_{1,\varepsilon}^{+/2} P \big(P' \widehat{\Sigma}_1^+(\varepsilon) P\big)^+ P' \big(\widehat{\Sigma}_{1,\varepsilon}^{+/2}\big)'.
$$
Then, the first-order-perturbation approximation can be calculated as
\begin{equation}\label{eqn:delta1}
\begin{aligned}
    \Delta_\varepsilon &= 
    c^2(n) \Vector (A_1)' (\widehat{Q}_{1,2}[\varepsilon] - \widehat{\mathcal{Q}}_{1,2}[\varepsilon]) \Vector(A_1) \\
    &=
    c^2(n) \Vector(A_1)' \big(\widehat{\Sigma}_1^{+/2}(\varepsilon)\big)' (\Delta \widehat{Q}_\varepsilon) \widehat{\Sigma}_1^{+/2}(\varepsilon) \Vector(A_1)\\
    &=
    2 c^2(n) \Vector(A_1)' \widehat{\Sigma}_1^{+}(\varepsilon) P_2 \big(P_2' \widehat{\Sigma}_1^+(\varepsilon) P_2\big)^+ (\Delta P_2)' (\widehat{\Sigma}_{1,\varepsilon}^{+/2} )' \times \\
    & \hspace{1cm} \big(I_{\widehat{r}} - \widehat{Q}_\varepsilon(P_2)\big) \widehat{\Sigma}_{1,\varepsilon}^{+/2} \Vector(A_1 - M_1)
    + o_{\mathcal{P}}\big(c^2(n) \Delta P_2\big)\\
    &= 2 c^2(n) \Vector(A_1)' \Sigma_1^{+} P_2 \big(P_2' \Sigma_1^+ P_2\big)^+ (\Delta P_2)' (\Sigma_{1}^{+/2} )' \times \\
    & \hspace{1cm} G \Sigma_1^{+/2} \Vector(A_1 - M_1)
    + o_{\mathcal{P}}\big(c^2(n) \Delta P_2\big)
\end{aligned}
\end{equation}
where $\Delta P_2 = \widehat{P}_2 - P_2$. Here in the last equality, we use the assumption that $\Sigma_1(\varepsilon) = \Sigma_1$, hence $\widehat{\Sigma}_1^+(\varepsilon) = \Sigma_1^+ + o_{\mathcal{P}}(1)$, $\widehat{\Sigma}_{1,\varepsilon}^{+/2} = \Sigma_1^{+/2} + o_{\mathcal{P}}(1)$, and $I_{\widehat{r}} - \widehat{Q}_\varepsilon(P_2) = G + o_{\mathcal{P}}(1)$.

From now on we omit the $o_{\mathcal{P}}\big(c^2(n) \Delta P_2\big)$ term of $\Delta_\varepsilon$ in \eqref{eqn:delta1} for the following proof since we are only interested in its stochastic limit.

Then, assume $G = U' U$ where $U \in \mathbb{R}^{r_{1,2} \times r}$, $U U' = I_{r_{1,2}}$, define
$$
\bm{w} = (\bm{y}', \bm{z}')', \hspace{0.45cm} \bm{y} := U \Sigma_1^{+/2} \Delta P_2 \big(P_2' \Sigma_1^+ P_2\big)^+ P_2' \Sigma_1^{+} \Vector(A_1), \hspace{0.45cm} \bm{z} := U \Sigma_1^{+/2} \Vector(A_1 - M_1).
$$

By \autoref{ass:normal}, $A_1$ converges in probability to the true matrix $M_1$, and $\widehat{P}_2$ is a consistent estimator of $P_2$ with rate $c(n)$, then according to delta method and Slutsky's theorem, we can define the matrix $\widecheck{\Sigma} = \widecheck{\Sigma}(M_1, M_2, \Sigma_1, \Sigma_2)$ in \autoref{lem:asyerror} to satisfy
$$
c(n) \bm{y} = c(n) U \Sigma_1^{+/2} \Delta P_2 \big(P_2' \Sigma_1^+ P_2\big)^+ P_2' \Sigma_1^{+} \Vector(A_1) \xrightarrow{\mathcal{D}} \mathcal{N}(0, \widecheck{\Sigma}).
$$
In addition, we have
$$
c(n) \bm{z} = c(n) U \Sigma_1^{+/2} \mbox{vec}(A_1 - M_1) \xrightarrow{\mathcal{D}} \mathcal{N}(0, I_{r_{1,2}}).
$$
And note that due to the independence between $A_1$ and $\Delta P_2$, the following expectation can be decomposed to
\begin{equation}\label{eqn:expProd}
    \mathbb{E}\Big[\big(c(n) \bm{z}\big) \big(c(n) \bm{y}\big)'\Big] = U \Sigma_1^{+/2} \mathbb{E} \bigg[ c(n) f(A_1) \bigg] \mathbb{E}\big[c(n) \Delta P_2' \big] (\Sigma_1^{+/2})' U',
\end{equation}

where $f(A_1) = \mbox{vec}(A_1 - M_1) \big(\Vector(A_1)\big)' \Sigma_1^{+} P_2 \big(P_2' \Sigma_1^+ P_2\big)^+$ is a matrix whose only randomness is from $A_1$. And since $c(n) \Delta P_2'$ has asymptotically zero mean, the expectation \eqref{eqn:expProd} is asymptotically zero, which indicates that $c(n) \bm{y}$ and $c(n) \bm{z}$ are asymptotically uncorrelated. Hence we have
$$
c(n) \bm{w} = c(n) \begin{pmatrix} \bm{y} \\ \bm{z} \end{pmatrix} \xrightarrow{\mathcal{D}} \mathcal{N}\Bigg(0, ~ \begin{pmatrix} \widecheck{\Sigma} & 0 \\ 0 & I_{r_{1,2}} \end{pmatrix}\Bigg),
$$
and consequently by the continuous mapping theorem,
$$
\Delta_\varepsilon = 2 c^2(n) \bm{w}' \begin{pmatrix} 0 & I_{r_{1,2}} \\ I_{r_{1,2}} & 0 \end{pmatrix} \bm{w} \xrightarrow{\mathcal{D}} 
Z, \hspace{0.12cm} \text{ with } \hspace{0.12cm} Z= 
\sum_{i = 1}^{r_{1,2}} 2\sqrt{\lambda_i} (\nu_{i,1} - \nu_{i,2}),
$$
where $\lambda_i$ are the eigenvalues of $\widecheck{\Sigma}$ and $\nu_{i,j} \stackrel{i.i.d.}{\sim} \chi^2(1)$ for $i \in \{1, \dots, r_{1,2}\}$, and $j = 1,2$. In addition, note that the variance of a $\chi^2(1)$ distributed random variable is 2. For this reason,
$$
\var(Z) = \sum_{i=1}^{r_{1,2}} 4 \lambda_i \big(\var(\nu_{i,1}) + \var(\nu_{i,2})\big) = 16 \sum_{i=1}^{r_{1,2}} \lambda_i = 16 \tr(\widecheck{\Sigma}).
$$
\end{proof}
\section{Generalized Wald test}\label{app:est}

In this section we provide a generic result to prove \Cref{thm:part.Est,,thm:multi_eig.Est,,thm:LLRest,,thm:comm.Est}. In fact, \autoref{thm:generalwald} below combined together with the asymptotic normality results proved in \Cref{thm:comm,,thm:LLR,,thm:multi_eig,,thm:part} gives \Cref{thm:part.Est,,thm:multi_eig.Est,,thm:LLRest,,thm:comm.Est}.

Consider an estimated vector $\widehat{\bm{v}}$ that depends on the sample size $n$ and satisfies the asymptotic normality statement 
\begin{equation}\label{eq:wald.asym}
    c(n) (\widehat{\bm{v}} - \bm{v}) \xrightarrow{\mathcal{D}} \mathcal{N}(0, \Sigma)
\end{equation}
with deterministic $\bm{v}$ and positive semidefinite limiting covariance matrix $\Sigma \in \mathbb{R}^{\tau \times \tau}$. Then, a Wald-type test statistic can be defined as
\begin{equation}\label{eq:Wald.eg}
    \Gamma = c^2(n) (\widehat{\bm{v}} - \bm{v})' \Sigma^+ (\widehat{\bm{v}} - \bm{v}) \xrightarrow{\mathcal{D}} \chi^2(\rank(\Sigma)).
\end{equation}
If $\Sigma$ is singular, the continuous mapping theorem is no longer applicable to justify replacing $\Sigma$ by a consistent estimator $\widehat{\Sigma}$ in \eqref{eq:Wald.eg}. In order to achieve a convergence result for $\Gamma$ in \eqref{eq:Wald.eg} based on an estimator for $\Sigma$, we introduce the following theorem.

\begin{thm}\label{thm:generalwald}
Assume \eqref{eq:wald.asym} holds and $\widehat{\Sigma}$ is a consistent estimator of $\Sigma$, threshold $\varepsilon > 0$ is not an eigenvalue of $\Sigma$. Then, alternative to \eqref{eq:Wald.eg}, we propose statistics with truncated SVD introduced in \autoref{subsec:Cov},
\begin{equation}\label{eq:wald.new.exact}
    \Gamma(\varepsilon) = c^2(n) (\widehat{\bm{v}} - \bm{v})' \Sigma^+(\varepsilon) (\widehat{\bm{v}} - \bm{v}) \xrightarrow{\mathcal{D}} \chi^2\big(\rank(\Sigma; \varepsilon)\big),
\end{equation}
and
\begin{equation}\label{eq:wald.new.est}
    \Gamma^\#(\varepsilon) = c^2(n) (\widehat{\bm{v}} - \bm{v})' \widehat{\Sigma}^+(\varepsilon) (\widehat{\bm{v}} - \bm{v}) \xrightarrow{\mathcal{D}} \chi^2\big(\rank(\widehat{\Sigma}; \varepsilon)\big).
\end{equation}
\end{thm}

Indeed if the threshold $\varepsilon$ is less than the smallest non-zero eigenvalue of $\Sigma$, especially when at the extreme case $\varepsilon = 0$, the limiting distribution in \eqref{eq:wald.new.exact} is exactly the same as \eqref{eq:Wald.eg}.

\begin{proof}[Proof of \autoref{thm:generalwald}]
Assume SVD gives $\Sigma = Z \Pi Z'$ with orthogonal $Z$, then by definition of the truncated SVD,
$$
\Sigma \times \Sigma^+(\varepsilon) = \Sigma^+(\varepsilon) \times \Sigma = Z \begin{pmatrix} I_{\rank(\Sigma; \varepsilon)} & 0 \\ 0 & 0 \end{pmatrix} Z'
$$
is an idempotent matrix with rank $\rank(\Sigma; \varepsilon)$.
Then, \eqref{eq:wald.new.exact} follows immediately by the continuous mapping theorem. According to \eqref{eq:wald.cmt} in \autoref{lem:ginverse}, we can find consistent estimators for the generalized inverse as well as for the rank of $\Sigma$. \autoref{lem:ginverse}, \eqref{eq:wald.new.exact} and the continuous mapping theorem prove \eqref{eq:wald.new.est}.
\end{proof}
\end{appendix}

\begin{acks}[Acknowledgments]
The authors thank the anonymous Reviewers and the Associate editors from both \emph{Journal of American Statistical Association} and \emph{Annals of Statistics} for their comments which helped improving the paper. 
\end{acks}
\begin{funding}
The authors gratefully acknowledge financial support from a Xerox PARC Faculty Research Award, National Science Foundation Awards 1455172, 1934985, 1940124, and 1940276, USAID, and Cornell University Atkinson Center for a Sustainable Future.
This research was conducted with support from the Cornell University Center for Advanced Computing, which receives funding from Cornell University, the National Science Foundation, and members of its Partner Program.
\end{funding}

\begin{supplement}
\stitle{Testing Simultaneous Diagonalizability}
\sdescription{Available from the repository \href{https://github.com/XycYuchenXu/Testing-Simultaneous-Diagonalizability}{Testing Simultaneous Diagonalizability}, the supplementary material includes data, code and output for reproducing the analysis in the paper. In addition, an \lstinline{R} package \lstinline{eigTest} is developed specifically for the project and made publicly accessible at \href{https://github.com/XycYuchenXu/eigTest}{Github}.}
\end{supplement}


\bibliographystyle{imsart-nameyear} 
\bibliography{bibliography}       


\end{document}